\documentclass[journal]{IEEEtran}

\usepackage{booktabs}
\usepackage{multirow}
\usepackage{color}
\usepackage{amssymb}
\usepackage{balance}
\usepackage{bm}
\usepackage{cite}
\usepackage{soul}

\usepackage{subcaption}
\usepackage{color}
\usepackage{framed}
\ifCLASSINFOpdf
 \usepackage[pdftex]{graphicx}
 \graphicspath{{../pdf/}{../jpeg/}}
 \DeclareGraphicsExtensions{.pdf,.jpeg,.png,.jpg}
\else
\fi
\usepackage{amsmath}
\usepackage{array}
\begin{document}

\title{
Deep Joint Demosaicing and High Dynamic Range Imaging within a Single Shot
}

\author{Yilun~Xu$^{\dagger}$, Ziyang~Liu$^{\dag}$, Xingming~Wu$^{\ast}$, Weihai~Chen,~\IEEEmembership{Member,~IEEE,}
	\\Changyun~Wen,~\IEEEmembership{Fellow,~IEEE,}
	and~Zhengguo~Li,~\IEEEmembership{Senior Member,~IEEE}
\thanks{$\dagger$ Joint first authors: Yilun Xu and Ziyang Liu.
}
\thanks{$\ast$ Corresponding author: Xingming Wu.
}
\thanks{Yilun Xu, Ziyang Liu, Xingming Wu, and Weihai Chen are
with the School of Automation Science and Electrical Engineering,
Beihang University, Beijing, 100191 (e-mail: yilunxu\_buaa@163.com, by1703126@buaa.edu.cn, wxmbuaa@163.com, and whchen@buaa.edu.cn).}
\thanks{Changyun Wen is with the School of Electrical and Electronics Engineering,
Nanyang Technological University, Singapore. Email: ecywen@ntu.edu.sg.}
\thanks{Zhengguo Li is with SRO Department, Institute for Infocomm Research, 1
Fusionopolis Way, Singapore (email: ezgli@i2r.a-star.edu.sg).}
}

\maketitle

\begin{abstract}
		Spatially varying exposure (SVE) is a promising choice for high-dynamic-range (HDR) imaging (HDRI). The SVE-based HDRI, which is called single-shot HDRI, is an efficient solution to avoid ghosting artifacts. However, it is very challenging to restore a full-resolution HDR image from a real-world image with SVE because: 
		a) only one-third of pixels with varying exposures are captured by camera in a Bayer pattern, b) some of the captured pixels are over- and under-exposed. For the former challenge, a spatially varying convolution (SVC) is designed to process the Bayer images carried with varying exposures. For the latter one, an exposure-guidance method is proposed against the interference from over- and under-exposed pixels. Finally, a joint demosaicing and HDRI deep learning framework is formalized to include the two novel components and to realize an end-to-end single-shot HDRI. Experiments indicate that the proposed end-to-end framework avoids the problem of cumulative errors and surpasses the related state-of-the-art methods.
\end{abstract}


\IEEEpeerreviewmaketitle

\section{Introduction}
\label{sec:intro}
The dynamic range of a natural scene is usually much higher than that of a low-dynamic-range (LDR) image captured using a smartphone or a digital camera via a single shot. Considerable information from the real scene is lost in the LDR image. HDRI technology was introduced to address such a problem \cite{HDR-book,CVPR2020-deepHDR,SIGGRAPH2020-deepHDR,eccv2020-deghost}. HDRI has become one of the hottest topics in the fields of image processing and computer vision.

A popular method for HDRI is to sequentially capture multiple LDR images with varying exposures sequentially and then merge them into an HDR image \cite{tcsvt-mef,lizheng3,1wang2020,deepHDR}. This method is called exposure stacking, which is widely adopted in smartphones and digital cameras. Exposure stacking performs well in static scenes. However, moving objects could exist in the shooting scenes. which lead to unavoidable ghosting artifacts in the HDR image
 \cite{lizheng1,ghosthard1,ghosthard2,lizheng2}.

Many methods were proposed to eliminate ghosting artifacts in the HDR image, but a large amount of computation is often required and these methods may fail in scenes with very complex motion and extreme dynamic range. 

\begin{figure}[htb]
	\setlength{\abovecaptionskip}{0pt}
	\setlength{\belowcaptionskip}{0pt}
	\begin{center}
		\includegraphics[width=1\linewidth]{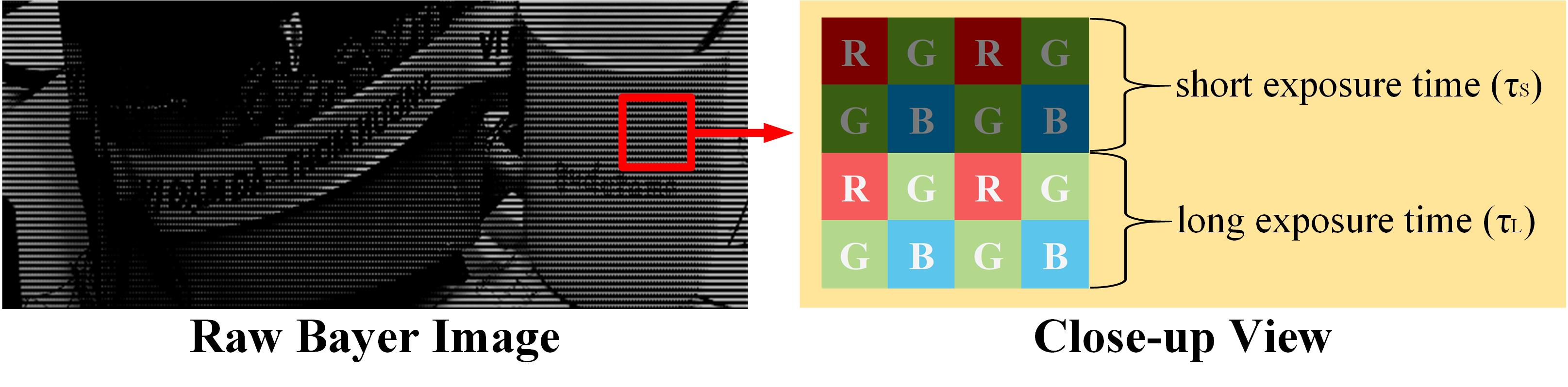}
	\end{center}
	\caption{Raw Bayer image with row-wise varying exposure times \cite{LeeASC}.}
	\label{fig:dualtime}
\end{figure}

Due to the challenge of ghosting removal \cite{ghosthard1,ghosthard2}, 
ghosting artifacts are believed to be the Achilles’ heel for the exposure stacking-based HDRI. As such, exposure stacking is unsuitable for capturing HDR videos \cite{1fran2016,1yang2016,1Wei2018}. The single-shot HDRI was proposed to capture ghost-free HDR images with varying exposures in a single image \cite{SVE,LeeASC,dualiso-TIP,ACCV}. In the single-shot HDRI, the exposure of pixels varies along with different spatial locations. 

In general, the raw data obtained by the camera is in a Bayer image. One typical example is given in Fig. \ref{fig:dualtime}, where the raw Bayer image is sensed by alternating the exposure time every other rows. This shooting method \cite{dual-time1, dual-time2, LeeASC} is referred to as dual-time in this paper. In addition, the single-shot HDRI is also a good candidate to capture HDR videos \cite{HDR-video-SVE}.

\begin{figure*}[htb]
	\setlength{\abovecaptionskip}{0pt}
	\setlength{\belowcaptionskip}{0pt}
	\begin{center}
		\includegraphics[width=1.0\linewidth]{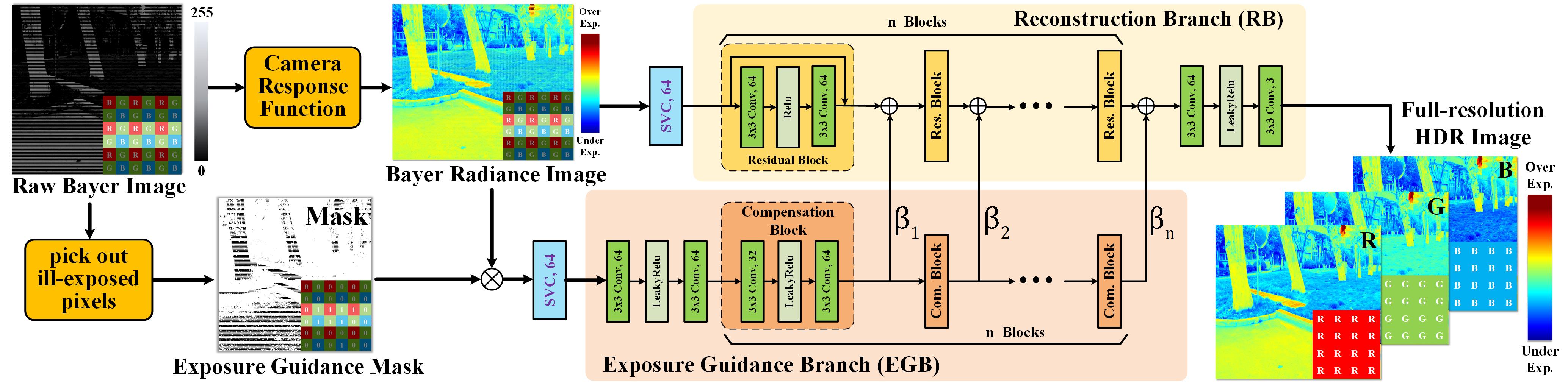}
	\end{center}
	\caption{Overall framework of the proposed algorithm. The proposed CNN includes a reconstruction branch (RB), and two distinctive components: spatially varying convolution (SVC) and exposure-guidance branch (EGB).}
	\label{fig:overall}
\end{figure*}

Among the single-shot dual-time HDRI algorithms \cite{dual-time1,dual-time2,LeeASC}, the stages of demosaicing and HDR reconstruction are separated. Thus, the error generated in the previous stage will affect the posterior one, resulting in error accumulation and drift. Recently, the idea of joint demosaicing with other low-level image processing tasks \cite{joint-fog, dm2016,joint-sr3,SRRI,joint-noise-sr} has appeared in some studies to avoid the cumulative error. However, the existing joint demosaicing algorithm's convolution methods have not been adjusted accordingly to the spatial change of the data pattern. Moreover, when the captured Bayer image with varying exposure times is converted into a Bayer radiance image \cite{dual-time2, LeeASC} by using camera response functions (CRFs) \cite{debevec1997}, the brightness difference caused by different exposure time can not be completely eliminated. The ill-exposed pixels in the Bayer radiance image will also interfere the convolutional neural network (CNN) \cite{LeeASC}.

In this paper, a one-stage CNN is proposed for the single-shot dual-time HDRI. As shown in Fig \ref{fig:overall}.This novel CNN can restore high-quality HDR image at the full resolution from a single Bayer radiance image with varying exposures, by which the problem of cumulative errors in \cite{LeeASC, dual-time2, dual-time1} can be avoided. Two distinctive components are introduced into the CNN to handle the challenges in single-shot HDRI. For the challenges of processing image pixels with varying colors (caused by the Bayer pattern) and varying exposures (caused by the dual-time), a novel spatially varying convolution (SVC) is designed and introduced as the first layer of CNN. The proposed SVC is adaptive to both varying colors and varying exposures, and can extract information from the dual-time Bayer image efficiently. The SVC can be easily inserted into other networks to process the Bayer image with or without SVE, and can be redesigned into other flexible variants when the data pattern or SVE changes. For the challenge of processing ill-exposed pixels, a novel exposure-guidance method is proposed, and the method is inspired by the dual-branch network \cite{SFT,GB}. By exposure-guidance method, the prior knowledge of ill-exposed pixels is exploited, and an exposure-guidance branch (EGB) is proposed to assist the CNN by incorporating this prior. 

Clearly, both the SVC and the exposure-guidance method can improve the explain-ability of the proposed one-stage CNN. Moreover, a new HDR dataset is proposed in this paper. The proposed dataset consists of 500 pairs of images with short and long exposures, respectively. The varying exposures are achieved by changing the exposure time. Images with camera shaking or object movement are filtered out artificially. Finally, extensive experimental results demonstrate the effectiveness of the proposed algorithm.

Overall, the four major contributions of this paper are as follows: 

\begin{itemize}
	\item A one-stage CNN with an improved explain-ability is proposed to address the problems of demosaicking and single-shot dual-time HDRI end-to-end..
	\item A novel SVC is introduced to process the Bayer image with or without SVE appropriately.
	\item An exposure-guidance method is proposed to reduce the interference of ill-exposed pixels.
	\item A new HDR dataset is proposed in this paper. Camera parameters for shooting are provided in detail for other related HDR researchers to use.
\end{itemize}

The rest of this paper is organized as follows. Relevant works are reviewed in Section \ref{review}. Details of the proposed algorithm are presented in Section \ref{algo}. Extensive experimental results are provided in Section \ref{experiment}. Lastly, concluding remarks are listed in Section \ref{conclusion}.

\section{Literature Review}
\label{review}
In this section, the relevant works on HDRI and demosaicing are reviewed.
\subsection{HDRI}
\subsubsection{stack-based HDRI}
The most popular method for generating an HDR image is to capture multiple differently exposed LDR images and merge all the LDR images into one HDR image. Such a method is called stack-based HDRI \cite{stackHDR-1,stackHDR-2,stackHDR-3,stackHDR-4}.  All the captured LDR images are firstly mapped into the corresponding radiance maps through the CRFs \cite{debevec1997}, and multiple radiance maps are then fused into an HDR image via a weighted average manner.

Exposure stacking-based methods often perform well in static scenes. 
Nevertheless, in dynamic scenes, the positions of moving objects in the exposure stack are different, resulting in ghost artifacts in HDR images. 
To remove the ghosting artifacts, one of the input images is selected as the reference image. All moving objects in other images are synchronized with those objects in the reference images. Ghost removal was widely studied, and many interesting algorithms were introduced 
\cite{ga1-1,ga2-1,ga3-1,lizheng1,lizheng2,eccv2020-deghost}.
However, when complex motion or extreme dynamic range occurs in the scene, all these algorithms could fail. As indicated in \cite{ghosthard1,ghosthard2}, no universal deghosting algorithm is available. Ghosting artifacts are thus believed to the Achilles' heel for the exposure stacking-based HDRI.

\subsubsection{Single-Shot HDRI}

An alternative solution is to obtain multiple exposure information of a scene via a single shot, and this solution is attractive for HDR videos \cite{HDR-video-SVE} and full light field reconstruction \cite{LightField}. Nayar and Suda proposed the concept of spatially varying exposure (SVE) in \cite{SVE,ACCV}, such that diverse pixel values of a single image are differently exposed. Two types of typical methods can be adopted to achieve SVE. 

One type is to change the ISO value at different positions of the sensor. Given an ISO-based SVE image, a few methods were proposed to restore the final HDR image, including the adaptive kernel regression-based method \cite{dual-iso4}, inpainting-based deinterlacing method \cite{dual-iso2}, joint learning-based method \cite{dual-iso5}, dictionary-based method \cite{cscode}, adaptive filter-based method \cite{dualiso-C++}, and deep learning-based method \cite{dualiso-TIP}. Nevertheless, the increase in ISO for a high exposure will amplify the camera noise, especially in low-lighting conditions. 

The other way to achieve SVE is to change exposure times \cite{LeeAccess,LeeASC,dual-time2,dual-time1}, in which higher-quality images than the ISO-based approach can be obtained, as indicated in \cite{debevec1997}. For example, an image can be shot with the exposure times varying every other lines, as shown in Fig. \ref{fig:dualtime}. Gu et al. proposed to adopt the structure of a coded rolling shutter as the readout structure of a CMOS image sensor \cite{dual-time1} and introduced several coding schemes and corresponding applications. Cho et al. designed a multistage processing flow to restore dual-time SVE Bayer images to HDR images gradually \cite{dual-time2}. An and Lee introduced an CNN-based technology to correct Bayer images, and then adopted demosaicing to obtain HDR images \cite{LeeASC}. 

The stages of HDR reconstruction and demosaicing are separated. Thus, the error generated in the previous stage will affect the posterior one, resulting in error accumulation and drift. In addition, to deal with the ill-exposed pixels in the Bayer radiance image, Cho \textit{et al.} \cite{dual-time2} completes the image by deleting ill-exposed pixels and then interpolating from neighbors. An \textit{et al.} \cite{LeeASC} takes the complete image as input and use CNN to correct ill-exposed pixels. Compared with \cite{dual-time2}, the CNN is allowed to make full use of the information in the Bayer radiance image \cite{LeeASC,LeeAccess}, but ill-exposed areas will also interfere with the imaging results of well-exposed areas during the calculation process.

\subsection{Image Demosaicing}
\label{Challenges of Demosaicing}
A color filter array (CFA) is put in front of a CMOS sensor to use a sensor designed for grayscale images and capture color images \cite{bayer}. Each pixel on a Bayer image taken in this way has only one of red, green, and blue. The image is subjected to a postprocessing algorithm called demosaicing, which is to complete the missing information in the Bayer image. 
In images without SVE, no brightness difference caused by the different exposures occurs. 
Existing demosaicing algorithms are divided into two categories: model-based \cite{dm1_pattern,tcsvt-dm3,tcsvt-dm1,tcsvt-dm2} and learning-based \cite{joint-fog, dm2016,joint-fog, joint-sr3, SRRI, joint-noise-sr}. 

To complete the missing two-thirds of pixels in the Bayer image, conventional model-based demosaicing algorithms \cite{dm1_pattern,tcsvt-dm3,tcsvt-dm1,tcsvt-dm2,dm5} usually adopt different interpolation schemes for different data patterns. In learning-based demosaicing \cite{joint-fog, dm2016,joint-sr3,SRRI,joint-noise-sr,dm-dp} the missing pixels are interpolated by CNN. However, existing deep learning-based algorithms share a contrary interpolation philosophy with conventional demosaicing algorithms. As indicated by \cite{dm-destroy}, the interpolators should be adaptive to the changing of data pattern. It is unreasonable to adopt a single same interpolation scheme for all data patterns, because missing colors that need to be interpolated vary with different data patterns. Thus, it is the same for deep learning, where different data patterns should be convolved by different convolution kernels, and same patterns by same convolution kernels. But their convolution method \cite{joint-fog, dm2016,joint-sr3,SRRI,joint-noise-sr} has not been adjusted accordingly due to the spatial change of the data pattern. 

Moreover, in the Bayer image with SVE, the CRFs cannot completely eliminate the brightness difference caused by different exposures times. Restoring an HDR image from a dual-time SVE Bayer image brings more challenges to demosaicing.

\section{Proposed Algorithm}
\label{algo}

Given a raw Bayer image captured within a single shot, the CRFs are firstly applied to generate the Bayer radiance image, and then the exposure-guidance mask is generated from a raw Bayer image. The Bayer radiance image is restored into an HDR radiance map via a novel CNN end-to-end, and the prior information in the exposure-guidance mask can assist the CNN. Then the CNN includes a reconstruction branch (RB), and two distinctive components: spatial varying convolution (SVC) and exposure-guidance method. Both the proposed SVC and exposure-guidance method make the CNN more explainable. The overall joint learning framework is shown in Fig. \ref{fig:overall}. It should be pointed out that the proposed algorithm is on top of our previous work \cite{ICME2021}.

\subsection{Generation of Bayer Radiance Image}

The raw input image $Z$ captured within a single shot is 
a N-bit (N can be 8, 10, 12, etc.)
Bayer image with row-wise varying exposure times \cite{LeeASC,dual-time2,dual-time1}, as shown in Fig. \ref{fig:dualtime}.
Let $\Delta t_{ij}$ be the exposure time of the pixel in the $i$th row, $j$th column of $Z$, then it is given as:
\begin{gather}
		\Delta t_{ij}=
		\begin{cases}
			{\tau_S,}&{i\bmod 4=1\; {\rm or}\; 2}\\
			{\tau_L,}&{i\bmod 4=3\; {\rm or}\; 0}
		\end{cases},\label{eq:dualtime}
\end{gather}
where $\tau_S$ and $\tau_L$ are the short and long exposure times, respectively.

\begin{figure}[htb]
	\setlength{\abovecaptionskip}{0pt}
	\setlength{\abovecaptionskip}{0pt}
	\begin{center}
		\includegraphics[width=1.0\linewidth]{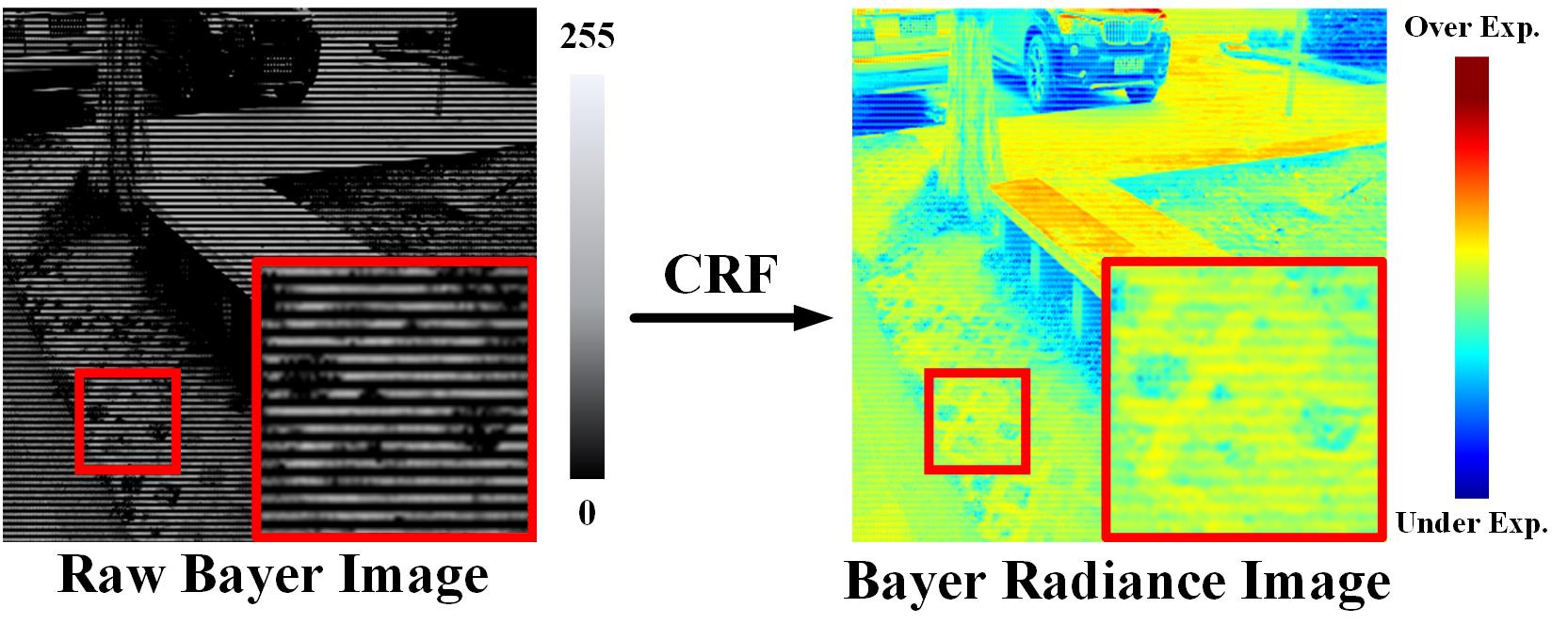}
	\end{center}
	\caption{Brightness differences are partly reduced using CRF. The slight horizontal stripes represent that the brightness differences cannot be completely eliminated.}
	\label{fig:difference}
\end{figure}

To restore the corresponding HDR image, the Bayer image $Z$ is firstly converted into a Bayer radiance image $E$ \cite{LeeASC}. 
The pixel $z_{ij}$ of $Z$ is normalized according to its exposure time.
\begin{gather}
		\ln (e_{ij}) = \ln (f_{c_{ij}}^{-1}(z_{ij})) - \ln(\Delta t_{ij}),
\end{gather}
where $c_{ij}\in\{R, G, B\}$ is the color channel of $z_{ij}$. The CRF $f_{c_{ij}}(\cdot)$ can be estimated via the method in \cite{debevec1997}. $f_{c_{ij}}^{-1}(\cdot)$ is the inverse function of $f_{c_{ij}}(\cdot)$. $e_{ij}$ represents the converted irradiance pixel. 
Note that CRFs are assumed to be available, because they can be easily estimated for any digital cameras and smartphones.
As shown in Fig. \ref{fig:difference}, the brightness differences caused by the dual-time are partially  reduced.

\begin{figure}[htb]
	\setlength{\abovecaptionskip}{0pt}
	\setlength{\belowcaptionskip}{0pt}
	\begin{center}
		\includegraphics[width=1.0\linewidth]{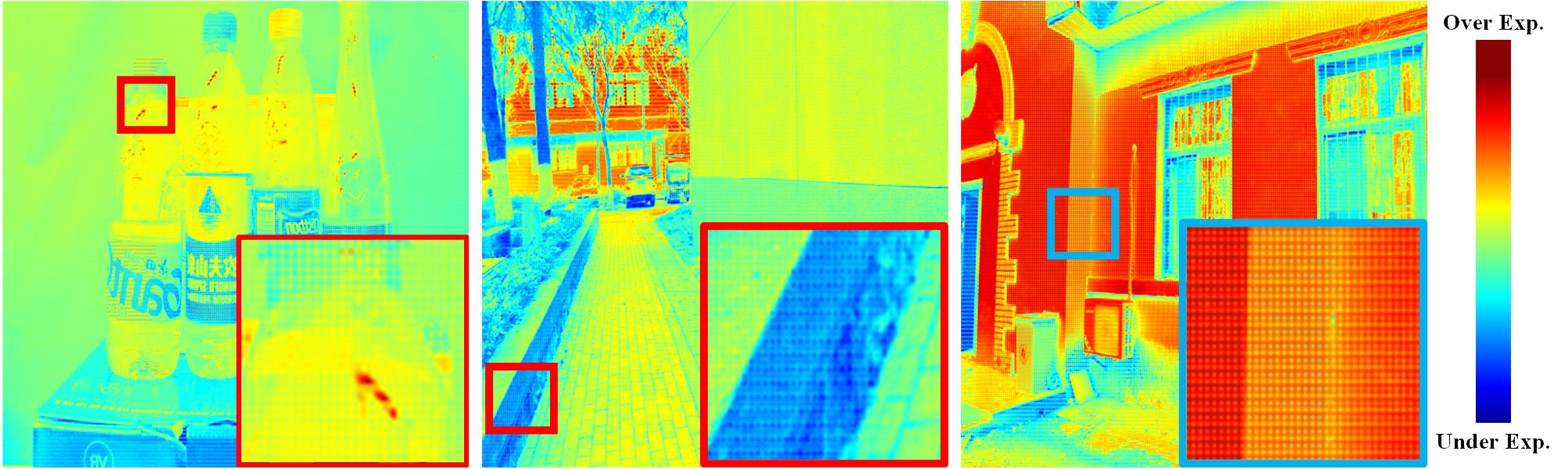}
	\end{center}
	\caption{Since only a single color (R, G, or B) is recorded at each pixel position, the visible grids appear in the Bayer radiance image.}
	\label{fig:mosaic}
\end{figure}

\begin{figure}[htb]
	\setlength{\abovecaptionskip}{0pt}
	\setlength{\abovecaptionskip}{0pt}
	\begin{center}
		\includegraphics[width=1\linewidth]{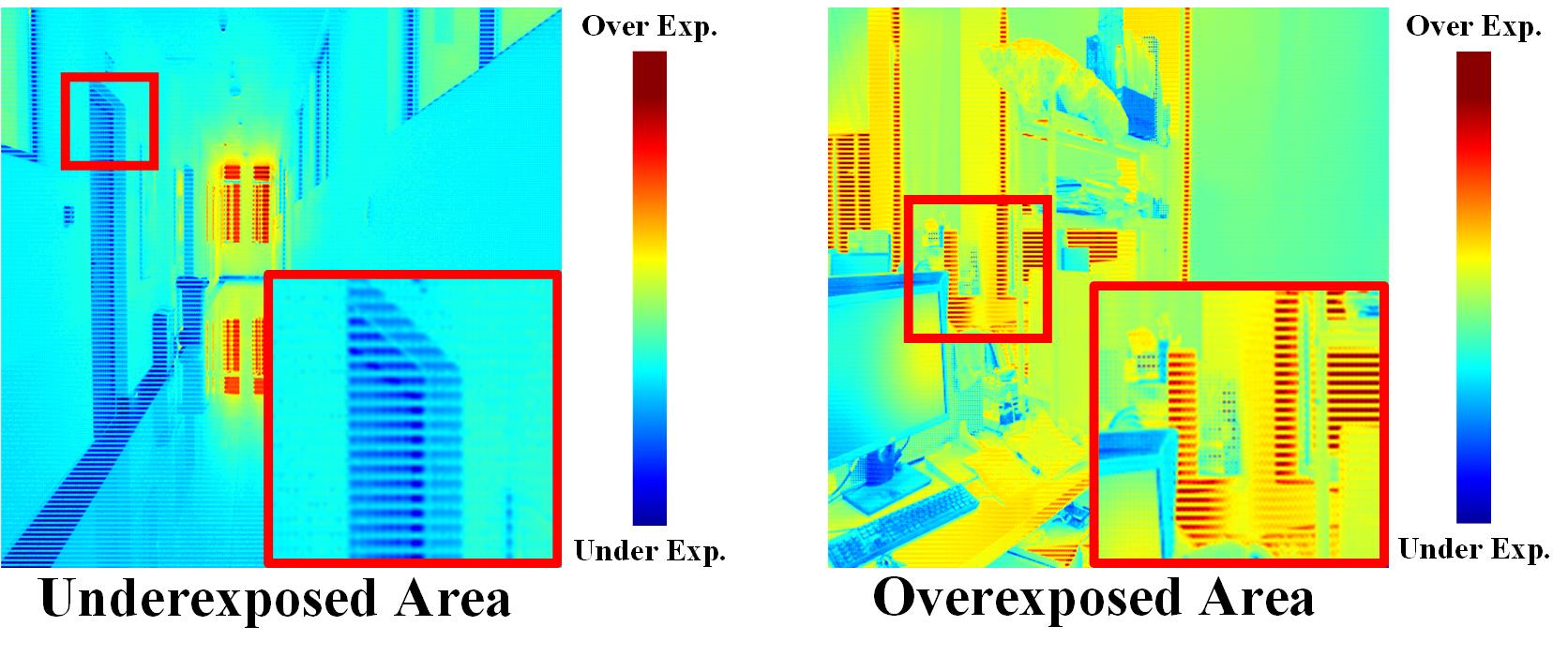}
	\end{center}
	\caption{Horizontal stripes indicating poor information, which is caused by over- and under-exposure.}
	\label{fig:error}
\end{figure}

As indicated above, there are three main challenges in restoring a full-resolution HDR image from a dual-time Bayer radiance image. First, two-thirds of color pixels are missed, resulting in visible grids, as shown in Fig. \ref{fig:mosaic}. Second, the remaining pixels are carried with varying exposures and the brightness differences cannot be completely eliminated via CRFs, like the horizontal stripes shown in Fig. \ref{fig:difference}. 
Moreover, some of the captured pixels are ill-exposed, which makes the HDR restoration more difficult, as shown in Fig. \ref{fig:error}.

\subsection{Spatially Varying Convolution}

\begin{figure}[htb]
	\setlength{\abovecaptionskip}{0pt}
	\setlength{\belowcaptionskip}{0pt}
	\begin{center}
		\includegraphics[width=1\linewidth]{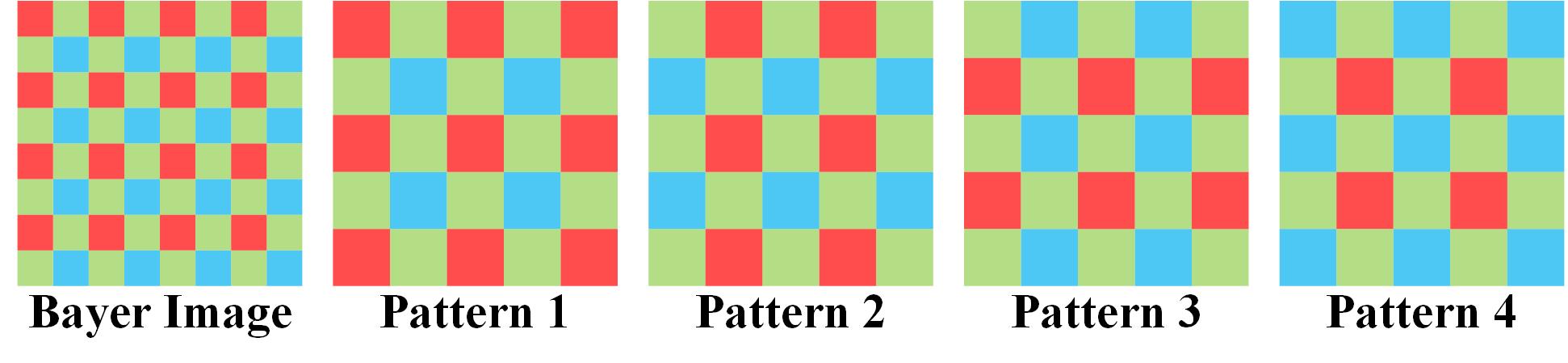}
	\end{center}
	\caption{Illustration of four different patterns in a $5 \times 5$ receptive field}
	\label{fig:4pattern}
\end{figure}

There are four color patterns in the Bayer image, as shown in Fig. \ref{fig:4pattern}. Conventional demosaicing methods complete pixels via color pattern-oriented interpolation schemes \cite{dm1_pattern,dm3_pattern,dm4_pattern}, which means that there exist different interpolators corresponding to the four different color patterns. However, demosaicing has not been adjusted accordingly across various color patterns in existing deep learning-based methods \cite{dm2016,joint-sr3,SRRI, joint-noise-sr}. All color patterns in an image are interpolated by a same kernel slidingly. It is difficult to realize adaptive interpolation due to weights sharing in the convolution.

\begin{figure}[htb]
	\setlength{\abovecaptionskip}{0pt}
	\setlength{\belowcaptionskip}{0pt}
	\begin{center}
		\includegraphics[width=0.9\linewidth]{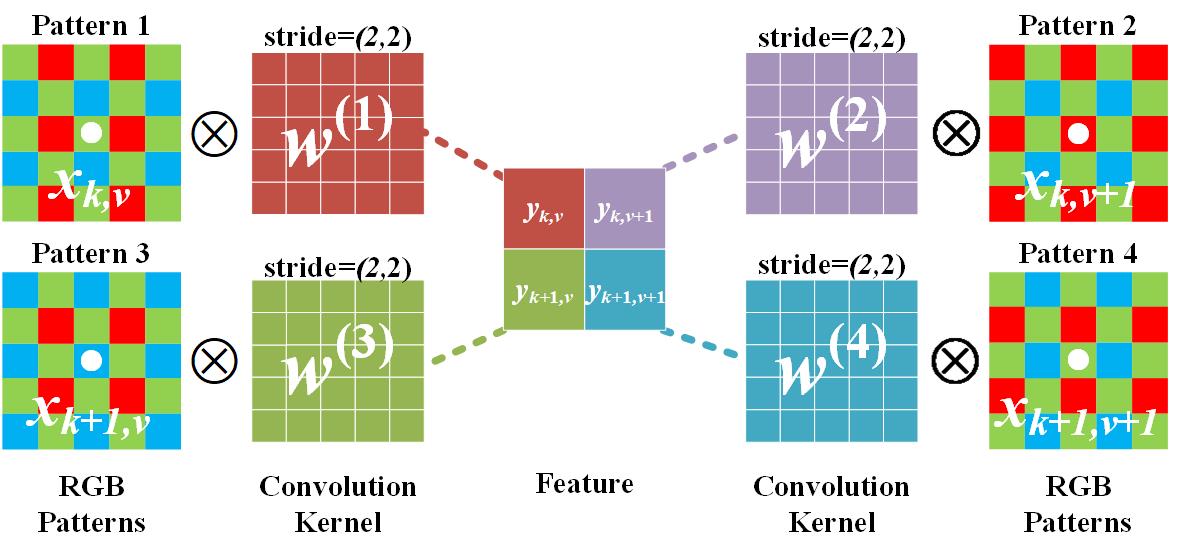}
	\end{center}
	\caption{Illustration of four different RGB patterns in a $5 \times 5$ receptive field, and the corresponding degraded version of spatially varying convolution (SVC-D).}
	\label{fig:svc4}
\end{figure}

We proposed a novel convolution method to use incomplete weight sharing for sliding kernels, which is called spatial varying convolution (SVC) \cite{ICME2021}. In the SVC, the kernel weights are shared across the same patterns but different across different patterns. Considering the four color patterns in Fig. \ref{fig:4pattern}, at least four different kernels need to be included in the SVC, as shown in Fig. \ref{fig:svc4}.

\begin{figure}[htb]
	\setlength{\abovecaptionskip}{0pt}
	\setlength{\belowcaptionskip}{0pt}
	\begin{center}
		\includegraphics[width=1\linewidth]{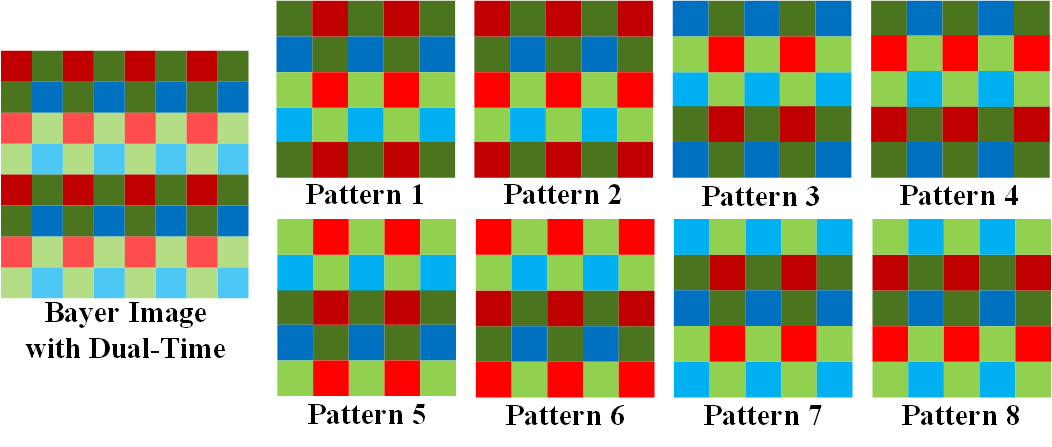}
	\end{center}
	\caption{Illustration of eight different patterns in a $5 \times 5$ receptive field, where the darker and lighter colors represent the short and long exposed radiation pixels, respectively.}
	\label{fig:8pattern}
\end{figure}

\begin{figure}[htb]
	\setlength{\abovecaptionskip}{0pt}
	\setlength{\belowcaptionskip}{0pt}
	\begin{center}
		\includegraphics[width=0.9\linewidth]{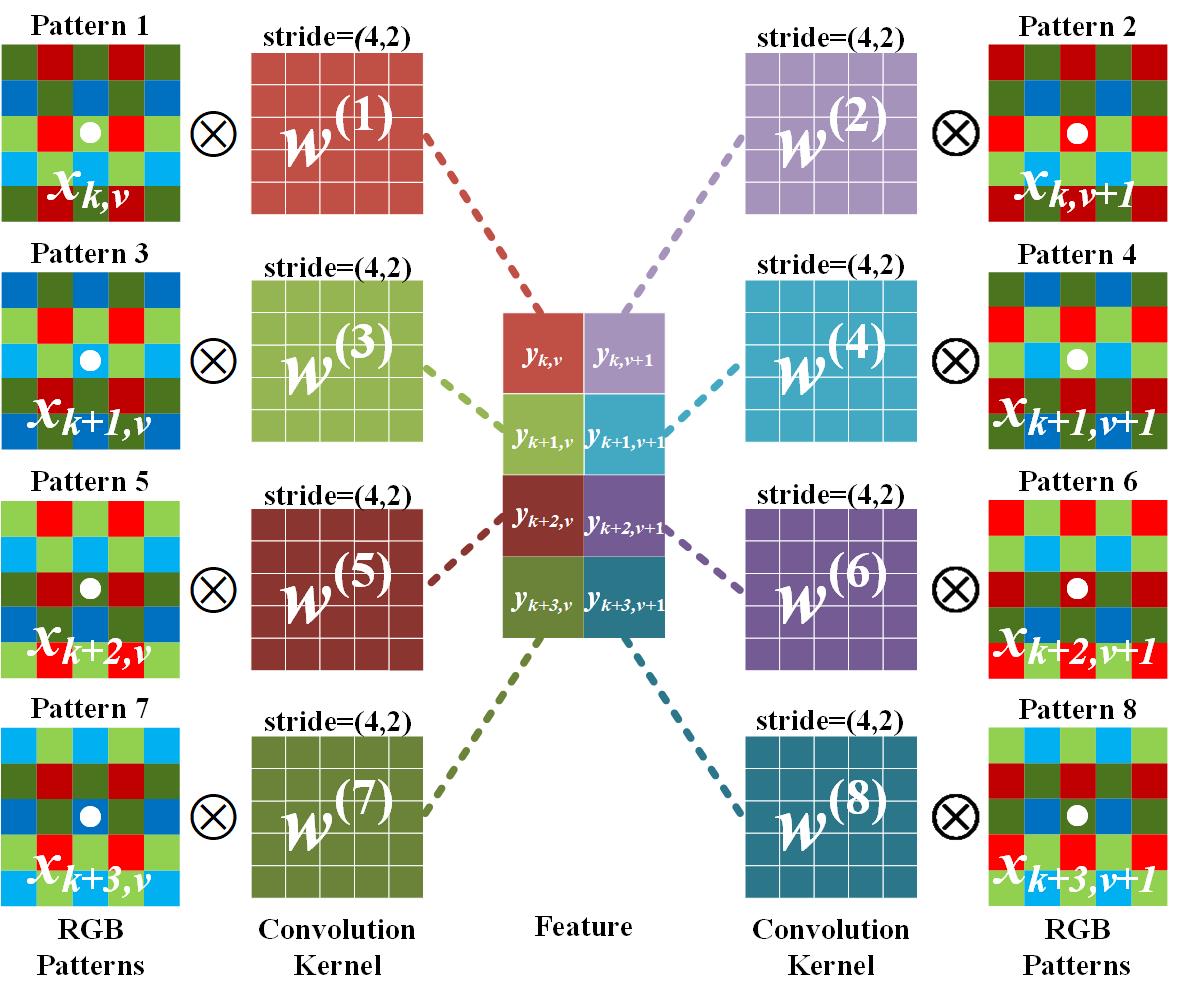}
	\end{center}
	\caption{Illustration of spatially varying convolution (SVC) based on 8 patterns.}
	\label{fig:svc8}
\end{figure}

Our proposed SVC \cite{ICME2021} is extended to a more complicated scenario. In a Bayer image captured by the dual-time shooting, the exposure time also varies along with image pixels. As shown in Fig. \ref{fig:difference}, the brightness differences, caused by the varying exposure times, cannot be eliminated by converting a Bayer image into a Bayer radiance image via the CRFs. Considering both varying colors and varying brightness, there are totally eight patterns in a $5 \times 5$ receptive field, as shown in Fig. \ref{fig:8pattern}. The darker and lighter colors represent the short and long exposed radiation values, respectively. Therefore, to adapt better to more kinds of patterns, the SVC is modified into a more complicated version, which includes eight interpolation kernels overall, as shown in Fig. \ref{fig:svc8}. The improved SVC is more robust to the varying patterns in a dual-time Bayer image. The details of the improved SVC are given in the following text.

Let $x_{k,v}$ and $y_{k,v}$ be the pixels at position $(k,v)$ in the input and output of SVC, respectively. $w^{(1)}$, $w^{(2)}$, $w^{(3)}$, $w^{(4)}$, $w^{(5)}$, $w^{(6)}$, $w^{(7)}$, and $w^{(8)}$ represent different convolution kernels. The proposed SVC is then defined as follows:
\begin{gather}\small{
		\begin{cases}
			{y_{k,v}}&=\sum\limits_{i=-2}^2 {\sum\limits_{j=-2}^2 {(w^{(1)}_{i,j} \times x_{k+i,v+j})} }\\
			{y_{k,v+1}}&=\sum\limits_{i=-2}^2 {\sum\limits_{j=-2}^2 {(w^{(2)}_{i,j} \times x_{k+i,v+1+j})} }\\
			{y_{k+1,v}}&=\sum\limits_{i=-2}^2 {\sum\limits_{j=-2}^2 {(w^{(3)}_{i,j} \times x_{k+1+i,v+j})} }\\
			{y_{k+1,v+1}}&=\sum\limits_{i=-2}^2 {\sum\limits_{j=-2}^2 {(w^{(4)}_{i,j} \times x_{k+1+i,v+1+j})} }\\
			{y_{k+2,v}}&=\sum\limits_{i=-2}^2 {\sum\limits_{j=-2}^2 {(w^{(5)}_{i,j} \times x_{k+2+i,v+j})} }\\
			{y_{k+2,v+1}}&=\sum\limits_{i=-2}^2 {\sum\limits_{j=-2}^2 {(w^{(6)}_{i,j} \times x_{k+2+i,v+1+j})} }\\
			{y_{k+3,v}}&=\sum\limits_{i=-2}^2 {\sum\limits_{j=-2}^2 {(w^{(7)}_{i,j} \times x_{k+3+i,v+j})} }\\
			{y_{k+3,v+1}}&=\sum\limits_{i=-2}^2 {\sum\limits_{j=-2}^2 {(w^{(8)}_{i,j} \times x_{k+3+i,v+1+j})} }\\
		\end{cases},\label{eq:svc8}	
}\end{gather}
where $k=0,4,8,...,4n_1\leq H$, and $v=0,2,4,...,2n_2\leq W$. $H$ and $W$ are the height and width of input or output, respectively. Benefiting from the SVC, the data pattern is the same for each convolution kernel, which can reduce the burden of network learning. {\it {\textbf {The proposed SVC is one distinctive component of the proposed CNN}}}.

\begin{figure}[htb]
	\setlength{\abovecaptionskip}{0pt}
	\setlength{\belowcaptionskip}{0pt}
	\begin{center}
		\includegraphics[width=1\linewidth]{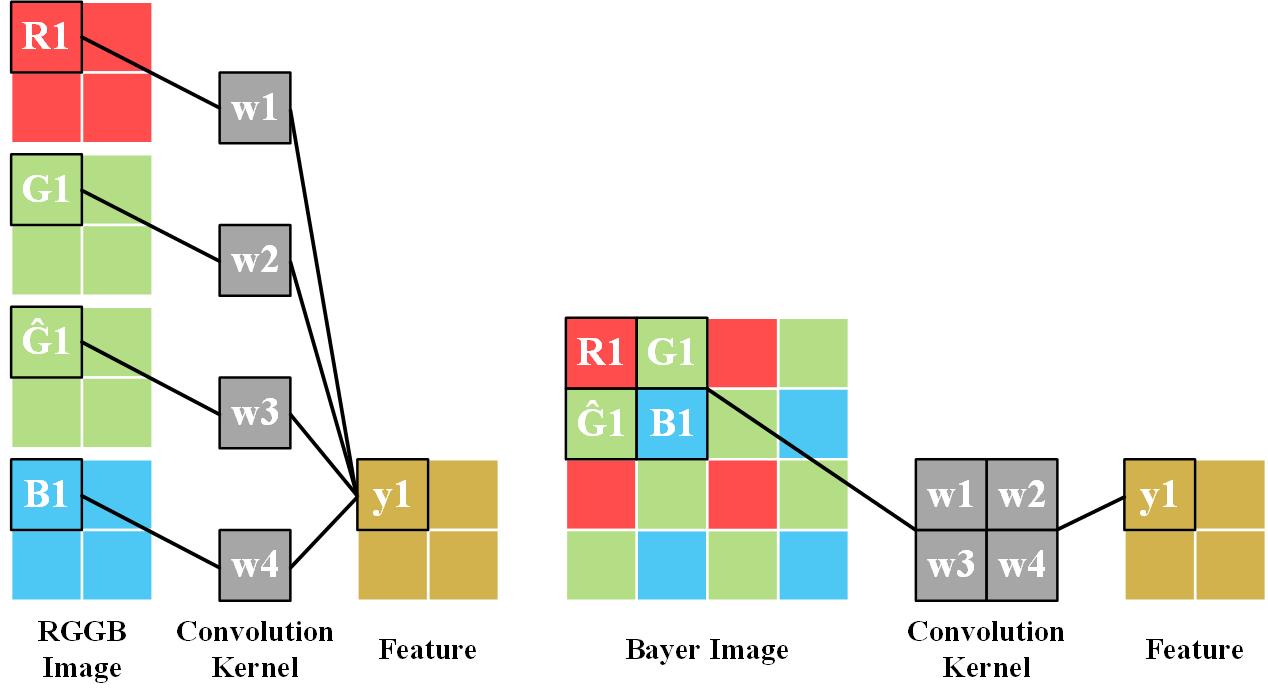}
	\end{center}
	\caption{Illustration of convolving an RGGB image which uses $1 \times 1$ convolution kernels at stride 1, and convolving a Bayer image which uses $2 \times 2$ convolution kernels at stride 2. The two operations are equivalent.}
	\label{fig:RGGB}
\end{figure}

Note that the difference between the SVC and existing methods is whether it is the convolution with complete weight sharing. Specifically, considering the algorithm in \cite{dm2016}, the Bayer image is convolved at stride 1. The convolutional kernel receives four kinds of patterns during sliding, as shown in Figure \ref{fig:4pattern}. According to the principle of convolution, kernel weights are shared, by which different data patterns are convolved by  the same convolutional kernels. Existing CNN-based demosaicing algorithms adopt two methods. One method is to set the stride as 2 or 4  \cite{joint-sr3}, by which the kernel receives only one kind of data pattern. Another method is to rearrange the single-channel Bayer image into a four-channel RGGB image, whose length and width are reduced by half. Then the rearranged image is convolved with a stride of 1 \cite{SRRI, joint-noise-sr}. The above two methods are actually equivalent. For example, convolving an RGGB image with a $1 \times 1$ kernel at stride 1 is equivalent to convolving a Bayer image with a $2 \times 2$ kernel at stride 2, as shown in Fig. \ref{fig:RGGB}. Both methods focus on interpolating only one kind of data pattern, where the other three kinds of patterns are not considered. While, the SVC can be adjusted accordingly to the spatial change of the data pattern.

Moreover the SVC can have flexible variants when the data pattern and SVE changes. The SVC is only used in the first layer of CNN, as shown in Fig. \ref{fig:overall}. Thus, compared to the weight-sharing convolution method, the SVC improves the imaging ability of CNN via minimal additional parameters.

\subsection{Reconstruction Branch}

The reconstruction branch (RB) shares a similar philosophy to the networks utilized in image-to-image translation tasks, such as image demosaicing \cite{dm-resnet}, super-resolution imaging \cite{sr-resnet}, and image denoising \cite{de-resnet}. The RB is composed of several residual blocks, as shown in Fig. \ref{fig:overall}. Each block is realized by an operation of convolution–ReLu–convolution and an identical mapping. Specifically, the kernel size in each convolutional layer is set to $3\times 3$. The stride and padding are both set to 1 to maintain the feature resolution during propagation. The structure of RB is shown in Fig. \ref{fig:overall}.

\subsection{Exposure-Guidance Method}

Over- and under-exposed areas exist in the rows with long- and short-exposure times, respectively. These ill-exposed areas are usually dominated by saturation noise. These areas lead to unreliable and poor information in the radiance map $E$, which turns into visible boundaries at the corresponding pixel positions, as shown in Fig. \ref{fig:error}. These ill-exposed areas can interfere with the CNN and are thus desired to be detected. To overcome this, an exposure-guidance method is proposed, which consists of exposure guidance mask and exposure guidance branch (EGB). {\it {\textbf {The exposure-guidance method is the other distinctive component of the proposed CNN}}}.

\subsubsection{Exposure-Guidance Mask}

In conventional HDRI algorithms 
\cite{lizheng1,lizheng2,tcsvt-mef,LeeASC}
, a threshold value is usually predefined on a N-bit image to identify the ill-exposed pixels. To simulate this artificial selection in a CNN-based algorithm, a exposure-guidance mask $M$ is computed using the following equation:

\begin{gather}
	\resizebox{.9\hsize}{!}{$
		m_{ij}=
		\begin{cases}
			{0,}&{\Delta t_{ij}=\tau_L\quad {\rm and}\quad z_{ij} \geq (1-\alpha)(2^N-1)}\\
			{0,}&{\Delta t_{ij}=\tau_S\quad {\rm and}\quad z_{ij} \leq \alpha(2^N-1) }\\
			{1,}&{otherwise}
		\end{cases},\label{eq:mask}
		$}
\end{gather}

where the $m_{ij}$ is the value at position $(i,j)$ in $M$. N is the number of bits for the input image and $\alpha$ is a percentage constant. Note that $M$ has the same size as $Z$ and $E$. Then, $M$ can be easily fed into the network along with the radiance map $E$, which can be considered as a prior of knowledge to help the network know which area is ill-exposed. Based on the rule of thumb in \cite{LeeASC} and the sensitivity analysis of $\alpha$ in Section \ref{Analysis of EGB}, threshold value $\alpha$ is empirically selected as $3.92\%$.

\subsubsection{Exposure-Guidance Branch}

The mask can be concatenated or element-wise multiplied with a Bayer radiance image as the network input, providing auxiliary prior information. However, the features of the radiance map are not guided by the mask in a deep-level manner because of this early fusion. Inspired by \cite{SFT,GB}, an EGB is proposed to guide the HDR reconstruction as an auxiliary branch, as shown in Fig. \ref{fig:overall}. Prior information can be extracted by this branch, which makes the CNN explainable.

During propagation, the exposure-guidance mask is first multiplied with the Bayer radiance image $E$. Thus, the poor information from the ill-exposed area will be filtered out, such that the EGB will tend to utilize the features extracted from the well-exposed area. Then, the features of the EGB are embedded into the RB in a multilevel manner to realize compensation, in which information is fused in a deeper level. Finally, the network can pay considerable attention to the well-exposed areas, which provide accurate HDR information.

The proposed EGB consists of $n$ blocks, and each block includes two convolutional layers and an activation layer. To reduce the computational cost, the output channels of the first convolutional layer are compressed in each block.

$E$ can be simply multiplied by $M$ as the input of RB to make the network focus on the well-exposed areas only. The experimental results in Section \ref{Analysis of EGB} show that this method does not work as expected. The possible reason is that the raw Bayer pattern is destroyed, and the irregular data makes the network difficult to deal with.

\subsection{Loss Function}

The $L_1$ loss function is widely used in deep learning-based low-level image processing \cite{L1loss}, and it is formalized as
\begin{gather}
		\mathcal{L}_{l1}=\frac{1}{3 \times H \times W}\sum\limits_{i = 1}^H{\sum\limits_{j=1}^W{|\boldsymbol {r}_{ij}-\tilde{\boldsymbol {r}}_{ij}|}},
\end{gather}
where $\boldsymbol {r}_{ij}$ and $\tilde{\boldsymbol {r}}_{ij}$ represent two 3D $(R, G, B)$ vectors 
at position $(i, j)$ in the image. The resolution of the ground truth $R$ and generated HDR image $\widetilde{R}$ is $H \times W$.

To reduce the color deviation between $R$ and $\widetilde{R}$, the color loss $\mathcal{L}_c$ is introduced as follows \cite{color-loss}:
\begin{equation}
		\mathcal{L}_c = \frac{1}{H \times W}\sum\limits_{i = 1}^H{\sum\limits_{j=1}^W{(1-cos(\boldsymbol {r}_{ij},\tilde{\boldsymbol {r}}_{ij}))}}, 
\end{equation}
where $cos(\boldsymbol {r}_{ij},\tilde{\boldsymbol {r}}_{ij})$ represents the cosine similarity of $\boldsymbol {r}_{ij}$ and $\tilde{\boldsymbol {r}}_{ij}$. $\mathcal{L}_c$ is sensitive to color difference. The overall loss function is given as

\begin{gather}
		\mathcal{L}_{total}=\mathcal{L}_{l1} + \lambda \mathcal{L}_c,
		\label{eq:loss}
\end{gather}
where $\lambda$ is a constant, and its value is selected as 0.1.

\section{Experimental Results}
\label{experiment}
\subsection{Implementation Details}

\subsubsection{Datasets Description}

Experiments are conducted on two datasets, VETHDR-Nikon dataset and VETHDR-Canon dataset, where the different exposures are achieved by varying the exposure times (VET) instead of changing the ISO. 
500 pairs of images are included in each dataset. Each pair consists of two full-resolution images $I_L$ and $I_S$, with long and short exposure times, respectively. The exposure time ratio and ISO are fixed as 16 and 800, during shooting respectively. All the images are resized to $480 \times 480$. Among each of the dataset, 300 pairs are used for training, 100 pairs for validation and 100 pairs for test. To simulate the input dual-time Bayer image $Z$, the pixels on every two rows are alternatively sampled from $I_S$ and $I_L$. The ground-truth HDR image $Y$ is obtained by merging the full-resolution images $I_S$ and $I_L$ via the method of \cite{debevec1997}.

The VETHDR-Nikon dataset is from \cite{chaobing,1zheng2021}, collected by a Nikon 7200 camera. It consists of original images in 8-bit color JPEG files format.
The corresponding CRF \cite{debevec1997} is recorded in the dataset. The Bayer Radiance Image can be calculated from the raw input image $Z$ through the nonlinear CRF.

The VETHDR-Canon dataset is collected by a Canon 5D4 camera in this paper. It contains 16-bit color images with the original digital counts for each of the RGB channels, which are generated from the 16-bit RAW image files by the method in \cite{raw-gt}. Since the RAW files typically have a nearly linear CRF, the Bayer Radiance Image can be calculated from the raw input image $Z$ through a simple linear transformation.

\subsubsection{Comparison Description}
In terms of qualitative comparison, the results are visualized by sequentially executing the tone mapping algorithm \cite{tonemap} and the white balance algorithm \cite{AWB} on the radiance map. In terms of quantitative comparison, we choose HDR-MAE, HDR-MSE, HDR-VDP, HDR-PSNR-RGB, HDR-SSIM-RGB, HDR-PSNR-Y, and HDR-SSIM-Y as the evaluation metrics. These metrics are measured on the radiance maps.
For the HDR-MAE and HDR-MSE, lower is better. For the HDR-VDP, HDR-PSNR-RGB, HDR-SSIM-RGB, HDR-PSNR-Y, and HDR-SSIM-Y, higher is better. Among them, a perceptually uniform (PU) encoding \cite{for-HDR} is utilzed to enable PSNR and SSIM to be used to evaluate the quality of HDR images. In addition, the model parameters (P) and the floating-point operations (FLOPs) are provided as reference. All evaluation metrics are calculated on the HDR radiance image, and which are explained as follows:
\begin{itemize}
	\item \textbf{HDR-MAE}: The average absolute value based on the PU encoding, which calculates the difference between the test image and the reference image.
	\item \textbf{HDR-MSE}: The average square value based on the PU encoding, which calculates the difference between the test image and the reference image.
	\item \textbf{HDR-VDP}: A high dynamic range visible difference predictor in version 2.2.2. Its quality correlate score can be used to evaluate image differences \cite{HDR-VDP}.
	\item \textbf{HDR-PSNR-RGB}: Peak signal-to-noise ratio based on the PU encoding, which calculates the difference of the image in the R, G, and B channels.
	\item \textbf{HDR-SSIM-RGB}: Structural similarity \cite{ssim} based on the PU encoding, which calculates the difference of the image in the R, G, and B channels.
	\item \textbf{HDR-PSNR-Y}: Peak signal-to-noise ratio based on the PU encoding, which calculates the difference of the image in the Y channel. 
	\item \textbf{HDR-SSIM-Y}: Structural similarity \cite{ssim} based on the PU encoding, which calculates the difference of the image in the Y channel. 
	\item \textbf{P}: Number of parameters in CNN.
	\item \textbf{FLOPs}: The number of floating-point operations required for network to generate an image.
\end{itemize}

\subsubsection{Training Details}
The RB and EGB are realized using 16 blocks. We randomly sample $128 \times 128$ patch from each input during training. We set the batch size and the number of iterations to 16 and $2 \times 10^5$, respectively. We set $\lambda$ in Equation (\ref{eq:loss}) to 0.1. The Adam optimizer with $\beta_1=0.9$ and $\beta_2=0.99$ is used for optimization. The learning rate is initially set as $2 \times 10^{-4}$ and finally decreased to $1 \times 10^{-7}$ through a cosine annealing schedule. Each model is trained on an NVIDIA GTX 1080Ti GPU for approximately two days. All the experiments are implemented using PyTorch.

\subsection{Analysis of SVC}
To demonstrate the effectiveness of the SVC, we compare the SVC with many other alternative methods qualitatively and quantitatively. We also perform a quantitative comparison of SVCs with different convolution kernel sizes.
\label{SVC_exp}

\subsubsection{Alternative Methods}

\begin{table*}[htbp]
	\centering
	\caption{Summary of various convolutional layers for Bayer images in existing algorithms.
		\label{tab:opt}
	}
	\renewcommand{\arraystretch}{1.4}
	\begin{tabular}{|c|c|ccc|c|}
		\hline
		\textbf{Method} & \textbf{Opt-base} & \textbf{Opt-2-2} \cite{dm2016} & \textbf{Opt-4-2} \cite{joint-sr3} & \textbf{Opt-4-4} \cite{joint-sr3} & \textbf{Opt-RGGB} \cite{SRRI, joint-noise-sr} \\ \hline
		\textbf{Input Shape} &  $h \times w \times 1$ &  $h \times w \times 1$ &  $h \times w \times 1$ &  $h \times w \times 1$ & $\frac{h}{2} \times \frac{w}{2} \times 4$ \\ \hline
		\textbf{Kernel Size} & $3 \times 3$ & $2 \times 2$ & $4 \times 4$ & $4 \times 4$ & $3 \times 3$ \\ \hline
		\textbf{Stride} & (1,1) & (2,2) & (2,2) & (4,4) & (1,1) \\ \hline
		\textbf{Upsampling Layer}  & Unused & 2 & 2 & 4 & 2 \\ \hline
		\textbf{Output Shape} &    $h \times w \times 64$ &   $h \times w \times 64$ &   $h \times w \times 64$ &   $h \times w \times 64$ &   $h \times w \times 64$ \\ \hline
	\end{tabular}
\end{table*}

The SVC is a flexible solution for processing a dual-time Bayer image, which ensures that pixels of different color patterns are convolved using different kernels. When the varying exposures caused by the dual-time are not considered, the SVC can be degraded to SVC-D \cite{ICME2021}. We further increase the kernel size of SVC-D to $5 \times 5$, as shown in Fig. \ref{fig:svc4}. The SVC-D is equivalent to adding the following conditions to Equation (\ref{eq:svc8}).
\begin{gather}
	\begin{cases}
		w^{(1)}_{i,j} = w^{(5)}_{i,j} &w^{(2)}_{i,j} = w^{(6)}_{i,j}\vspace{1ex}\\
		w^{(3)}_{i,j} = w^{(7)}_{i,j} &w^{(4)}_{i,j} = w^{(8)}_{i,j}\\
	\end{cases},\label{eq:svc4}
\end{gather}
In addition to the proposed SVC and SVC-D in this paper, a few special convolutional layers in deep learning have been utilized to process a special Bayer image in advance \cite{dm2016,joint-sr3,SRRI, joint-noise-sr}. All these layers/methods are summarized in Table \ref{tab:opt}. The first column indicates the method characteristics, which are explained as follows:
\begin{itemize}
	\item \textbf{Input Shape}: The shape of the input Bayer image, including the height, width, and number of channels.
	\item \textbf{Kernel Size}: The size of kernel in the first convolutional layer.
	\item \textbf{Stride}: The stride during kernel sliding in the first convolutional layer, including the vertical and horizontal strides. 
	\item \textbf{Upsampling Layer}: The upsampling layer is realized by the sub-pixel convolution \cite{sub-pixel}. 2 and 4 indicate the scaling factor.
	\item \textbf{Output Shape}: The shape of the output produced by the first convolutional layer, including the height, width and number of channels.
\end{itemize}

Table \ref{tab:opt} presents that two types of convolutional layers can be used for processing a Bayer image. One type is that a  convolution kernel of even size extracts the color information with a stride of even. Then, the downsampled output is resized to the original resolution by using the upsampling layer, such as Opt-2-2 \cite{dm2016}, Opt-4-2 \cite{joint-sr3}, Opt-4-4 \cite{joint-sr3}, and Opt-RGGB \cite{SRRI, joint-noise-sr}. Overall, all these convolutional layers are designed toward the same goal, i.e., to make the color pattern convolved using the kernel be the same. Opt-base represents the general convolutional layer with a kernel size of $3 \times 3$.

\begin{table*}[htbp]
	\setlength{\abovecaptionskip}{0pt}
	\setlength{\belowcaptionskip}{0pt}
	\centering
	\scriptsize
	\caption{Quantitative comparison among the baseline, existing operations, and our SVC. The best results are shown in bold, and the second-best results are shown in blue. The results come from the VETHDR-Nikon test set.
		\label{tab:SVC}
	}
	\resizebox{2.0\columnwidth}{!}{\begin{tabular}{c|ccccccccc}
			\toprule[1pt]
			\textbf{Method} & \textbf{HDR-MAE} & \textbf{HDR-MSE} & \textbf{HDR-VDP}& \textbf{HDR-PSNR-RGB}  & \textbf{HDR-SSIM-RGB}  & \textbf{HDR-PSNR-Y}  & \textbf{HDR-SSIM-Y}  & \textbf{P [$10^{6}$]} & \textbf{FLOPs [$10^{11}$]} \\
			\midrule
			\textbf{Opt-base} & {2.982} & { 53.874 } & { 65.59 } & { 41.38 } & { 0.9786 } & { 43.99 } & { 0.9860 } & \textbf{ 1.221 } & {\color{blue}2.813} \\
			\midrule
			\textbf{Opt-2-2} & { 2.911 } & { 41.480 } & {\color{blue}65.72} & { 41.72 } & { 0.9787 } & { 44.12 } & { 0.9859 } & {\color{blue}1.222} & \textbf{ 2.813 } \\
			\textbf{Opt-4-2} & { 2.962 } & { 42.596 } & { 65.55 } & { 41.61 } & { 0.9782 } & { 44.02 } & { 0.9856 } & { 1.225 } & { 2.814 } \\
			\textbf{Opt-4-4} & { 2.983 } & { 41.916 } & { 65.57 } & { 41.62 } & { 0.9783 } & { 44.00 } & { 0.9856 } & { 1.238 } & { 2.814 } \\
			\textbf{Opt-RGGB} & {2.903} & { 41.159 } & { 65.68 } & { 41.79 } & { 0.9788 } & { 44.20 } & { 0.9859 } & { 1.230 } & { 2.817 } \\
			\midrule
			\textbf{SVC-D} & {\color{blue}2.889} & {\color{blue}40.744} & \textbf{65.74} & {\color{blue}41.84} & {\color{blue}0.9790} & {\color{blue}44.21} & {\color{blue}0.9860} & { 1.227 } & { 2.816 } \\
			\textbf{SVC} & \textbf{ 2.881 } & \textbf{ 40.052 } & { 65.69 } & \textbf{41.91} & \textbf{0.9791} & \textbf{44.33} & \textbf{0.9861} & { 1.234 } & { 2.816 } \\
			\bottomrule[1pt]
	\end{tabular}}
\end{table*}

The convolutional layers in Table \ref{tab:opt} are combined with the RB to compare their performance. The results of 7 evaluation metrics on the radiance map are reported in Table \ref{tab:SVC}. The number of parameters is also provided as reference. The SVC and SVC-D surpass other convolutional layers, which demonstrates that the network learning can be affected by the varying color patterns and the SVC is a more robust solution. Moreover, the leading performance of SVC over SVC-D can prove the necessity of further consideration of brightness difference in a dual-time Bayer image.

\begin{figure*}[htb]
	\setlength{\abovecaptionskip}{0pt}
	\setlength{\belowcaptionskip}{0pt}
	\begin{center}
		\includegraphics[width=1.0\linewidth]{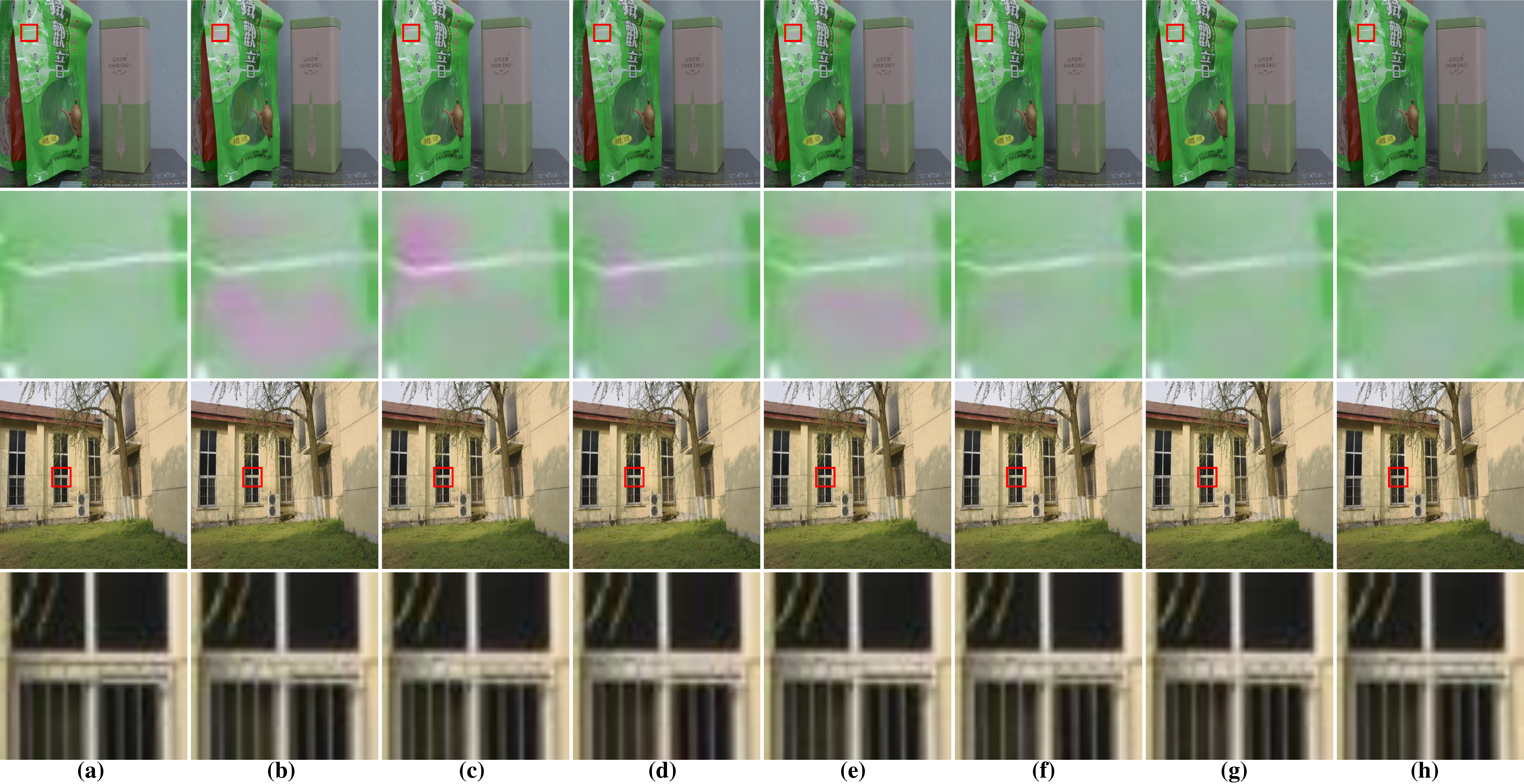}
	\end{center}
	\caption{Qualitative comparison among the baseline, existing convolutional layers, and our SVC. (a) Ground truth, (b) Opt-base, (c) Opt-2-2, (d) Opt-4-2, (e) Opt-4-4, (f) Opt-RGGB, (g) SVC-D, (h) SVC. The results come from the VETHDR-Nikon test set.}
	\label{fig:SVC_exp}
\end{figure*}

The qualitative results are shown in Fig. \ref{fig:SVC_exp}. In the first row, \textbf{(b)}, \textbf{(c)}, \textbf{(d)}, \textbf{(e)}, and \textbf{(f)} fail in estimating correct colors. In the second row, clearer edges are recovered using \textbf{(h)}, whereas other results have different degrees of blurring on edges. The SVC can make the network restore correct colors and textures.

\subsubsection{Kernel Size}

\begin{table*}[htbp]
	\setlength{\abovecaptionskip}{0pt}
	\setlength{\belowcaptionskip}{0pt}
	\centering
	\scriptsize
	\caption{Quantitative comparison among SVC-3, SVC-5, and SVC-7. The best results are shown in bold, and the second-best results are shown in blue. The results come from the VETHDR-Nikon test set.
		\label{tab:SVC_size}
	}
	\resizebox{2.0\columnwidth}{!}{\begin{tabular}{c|ccccccccc}
			\toprule[1pt]
			\textbf{Method} & \textbf{HDR-MAE} & \textbf{HDR-MSE} & \textbf{HDR-VDP}& \textbf{HDR-PSNR-RGB}  & \textbf{HDR-SSIM-RGB}  & \textbf{HDR-PSNR-Y}  & \textbf{HDR-SSIM-Y}  & \textbf{P [$10^{6}$]} & \textbf{FLOPs [$10^{11}$]}\\
			\midrule
			\textbf{Baseline} & {2.982} & {53.874} & {65.59} & {41.38} & {0.9786} & {43.99} & {0.9860} & \textbf{ 1.221 } & \textbf{2.813}\\
			\midrule
			\textbf{SVC-3} & {2.941} & {40.809} & \textbf{65.70} & { 41.80 } & { 0.9788 } & { 44.19 } & { 0.9859 } & {\color{blue}1.225} & \textbf{2.813} \\
			\textbf{SVC-5} & \textbf{2.881} & \textbf{40.052} & {\color{blue}65.69} & \textbf{41.91} & \textbf{0.9791} & \textbf{44.33} & \textbf{0.9861} & { 1.234 } & { \color{blue} 2.816 } \\
			\textbf{SVC-7} & {\color{blue}2.902} & {\color{blue}40.548} & { 65.68 } & {\color{blue}41.87} & {\color{blue}0.9790} & {\color{blue}44.28} & {\color{blue}0.9860} & { 1.246 } & { 2.819 } \\
			\bottomrule[1pt]
	\end{tabular}}
\end{table*}

The SVCs with different kernel sizes are also experimented. In Table \ref{tab:SVC_size}, SVC-3, SVC-5, and SVC-7 indicate a kernel size of $3 \times 3$, $5 \times 5$, and $7 \times 7$, respectively. All the SVCs outperform  the original convolution (Opt-base). It can be seen that simply increasing the kernel size of SVC brings little gains on performance. Thus, the power of SVC lies in the design philosophy, not the increasing parameters.

\subsection{Analysis of Exposure Guidance}
\label{Analysis of EGB}

To demonstrate the effectiveness of the exposure-guidance method, we compare the exposure-guidance method with many other alternative methods qualitatively and quantitatively. We also perform a quantitative comparison of EGBs with different threshold values $\alpha$. Moreover, we analyze the working mechanism of EGB through feature map and $\beta_i$.

\subsubsection{Alternative Methods}

The exposure-guidance mask is necessary to incorporate into the main network RB by adopting the EGB. Besides the EGB, there are three other ways in incorporating the exposure-guidance mask. The first method is to multiply the exposure-guidance mask and the Bayer radiance image to obtain an input matrix with a size of $h \times w \times 1$, in which the value of ill-exposed pixels becomes 0. This matrix is directly fed into the main network RB. The second method is to concatenate the exposure-guidance mask with the Bayer radiance image to generate a $h \times w \times 2$ matrix, which is then fed into the RB. The third method is to use the exposedness-aware compensation branch (EACB) to achieve deep-level fusion of masks and features \cite{ICME2021}. To make a fair comparison, for the first two method, we set the number of residual blocks in the RB as 25. For the third method and the EGB, we set both the number of blocks in RB and EGB as 16.

\begin{table*}[htbp]
	\setlength{\abovecaptionskip}{0pt}
	\setlength{\belowcaptionskip}{0pt}
	\centering
	\scriptsize
	\caption{
		Quantitative comparison between the baseline and four different methods of using exposure-guidance masks. The best results are shown in bold, and the second-best results are shown in blue. The results come from the VETHDR-Nikon test set.
		\label{tab:EGB_exp}
	}
	\resizebox{2.0\columnwidth}{!}{\begin{tabular}{c|ccccccccc}
			\toprule[1pt]
			\textbf{Method} & \textbf{HDR-MAE} & \textbf{HDR-MSE} & \textbf{HDR-VDP}& \textbf{HDR-PSNR-RGB}  & \textbf{HDR-SSIM-RGB}  & \textbf{HDR-PSNR-Y}  & \textbf{HDR-SSIM-Y}  & \textbf{P [$10^{6}$]} & \textbf{FLOPs [$10^{11}$]}\\
			\midrule
			\textbf{Baseline} & {2.982} & {53.874} & {65.59} & {41.38} & {0.9786} & {43.99} & {0.9860} & \textbf{1.221} & \textbf{2.813}\\
			\midrule
			\textbf{Multiplication} & {3.314} & {87.445} & {64.74} & {40.42} & {0.9757} & {43.24} & {0.9843} & {1.886} & {4.345}\\
			\textbf{Concatenation} & {2.904} & {50.018} & {\color{blue}65.75} & {41.50} & {0.9790} & {44.11} & {0.9862} & {1.886} & {4.346}\\
			\midrule
			\textbf{RB+EACB \cite{ICME2021}} & {\color{blue}2.900} & {\color{blue}48.224} & {65.68} & {\color{blue}41.54} & {\color{blue}0.9790} & {\color{blue}44.12} & {0.9862} & {2.072} & {4.773}\\
			\textbf{RB+EGB} & \textbf{2.861} & \textbf{47.835} & \textbf{65.83} & \textbf{41.72} & \textbf{0.9794} & \textbf{44.32} & \textbf{0.9865} & {\color{blue}1.850} & {\color{blue}4.263}\\
			\bottomrule[1pt]
	\end{tabular}}
\end{table*}

The 7 evaluation metrics on the test set are reported in Table \ref{tab:EGB_exp}. Among them, the result of the multiplication-based method is worst. This implies that the information is not utilized after resetting the ill-exposed pixels to 0. The raw RGB pattern in the Bayer image is destroyed, and the irregular data make the network difficult to learn. The concatenation-based method is improved compared with the baseline on the 7 evaluation metrics, indicating that the awareness of ill-exposed pixels can improve the performance. However, the prior of the exposure-guidance mask is not incorporated into the network in a deep-level way, in which the performance improvements are limited. The EACB \cite{ICME2021} is able to incorporate the prior knowledge of ill-exposed pixels in the feature level. The proposed EGB is a modification of EACB and compensates the RB in a both multi-level and feature-level way. Since the EACB-based method \cite{ICME2021} only compensates RB once at the deepest level, the lack of compensation times make the performance improvement less obvious than the concatenation-based method. The method of using the EGB achieves the best results on all four indicators, which strongly proves the effectiveness of fusing the prior in a deep- and multi-level way. The EGB can focus only on the well-exposed area, and can exploit accurate information for HDR reconstruction. This network structure design is also more explainable. Moreover, the original information of the dual-time Bayer image is preserved. The EGB also has the highest computational efficiency.

\begin{figure*}[htb]
	\setlength{\abovecaptionskip}{0pt}
	\setlength{\belowcaptionskip}{0pt}
	\begin{center}
		\includegraphics[width=1\linewidth]{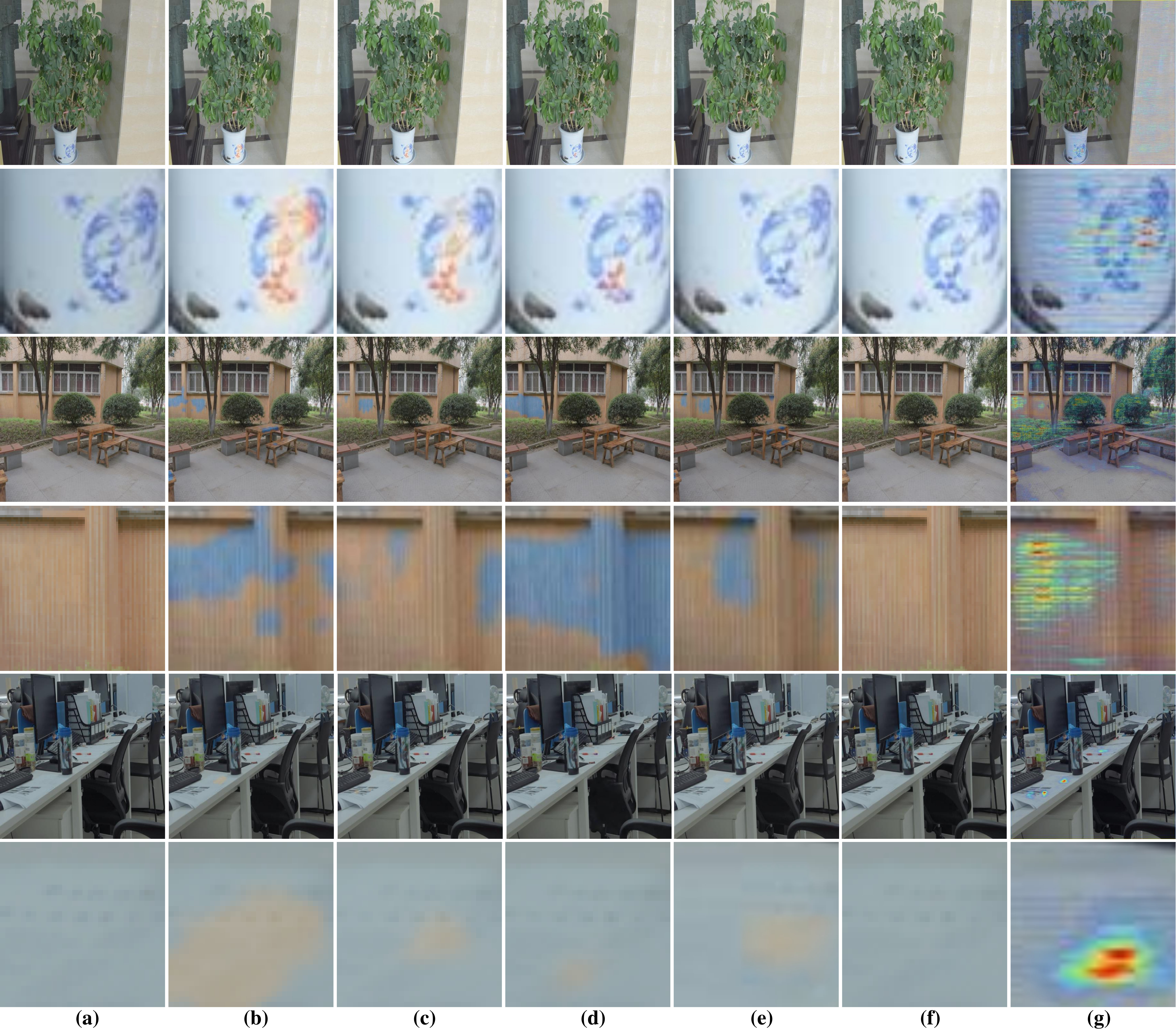}
	\end{center}
	\caption{Qualitative comparison between the baseline and three different methods of using exposure-guidance masks. (a) Ground truth, (b) baseline, (c) the method of using multiplication, (d) the method of using concatenation, (e) RB+EACB \cite{ICME2021}, (f) RB+EGB, (g) overlay of EGB's feature output and tone-mapped ground truth. The results come from the VETHDR-Nikon dataset.}
	\label{fig:EGB_exp}
\end{figure*}

The qualitative results are shown in Fig. \ref{fig:EGB_exp}. The EGB-based method performs evidently better on the four sets of images, whereas the other methods cause different degrees of unnatural color distortions.

\begin{table}[htbp]
	\setlength{\abovecaptionskip}{0pt}
	\setlength{\belowcaptionskip}{0pt}
	\centering
	\scriptsize
	\caption{$\beta$ in the trained RB+EGB.
		\label{tab:beta}
	}
	\begin{tabular}{cccccccc}
		\toprule[1pt]
		\bm{$\beta_1$} & \bm{$\beta_2$} & \bm{$\beta_3$} & \bm{$\beta_4$} & \bm{$\beta_5$}  & \bm{$\beta_6$}  & \bm{$\beta_7$}  & \bm{$\beta_8$}   \\
		\midrule
		0.6463 & 0.3757 & 0.6446 & 0.8103 & 0.7389 & 0.6234 & 0.4541 & 0.6906  \\
		\midrule
		\bm{$\beta_9$} & \bm{$\beta_{10}$} & \bm{$\beta_{11}$} & \bm{$\beta_{12}$} & \bm{$\beta_{13}$}  & \bm{$\beta_{14}$}  & \bm{$\beta_{15}$}  & \bm{$\beta_{16}$}   \\
		\midrule
		0.5762 & 0.6027 & 0.6895 & 0.7473 & 0.8965 & 0.9381 & 0.9597 & 0.9734  \\
		
		\bottomrule[1pt]
	\end{tabular}
\end{table}

\subsubsection{Threshold Selection}
\begin{figure}[htbp]
	\setlength{\abovecaptionskip}{0pt}
	\setlength{\belowcaptionskip}{0pt}
	\begin{center}
		\includegraphics[width=0.9\linewidth]{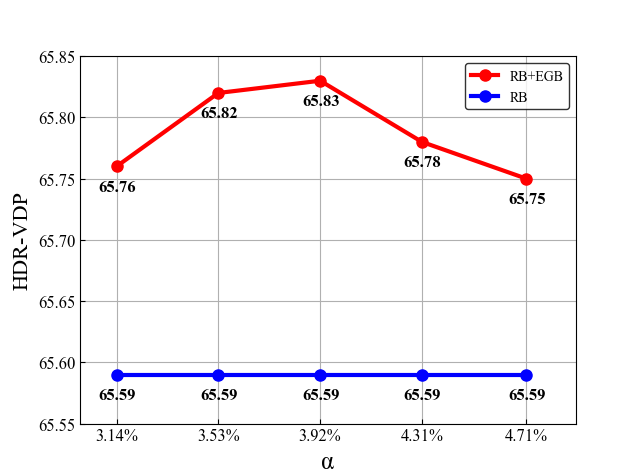}
	\end{center}
	\caption{The HDR-VDP score of the models under different $\alpha$. The results come from the VETHDR-Nikon test set.}
	\label{fig:zth}
\end{figure}

In order to observe the effect of different $\alpha$ on the EGB, we chose 5 values for testing. The selected values are all around the empirical value $3.92\%$ given in \cite{LeeASC}. The HDR-VDP scores of RB+EGB and RB under different $\alpha$ are shown in Figure \ref{fig:zth}. It can be observed from the figure that the performance of EGB is less affected by the changing of $\alpha$, and all the scores of RB+EGB significantly surpass the baseline RB. This proves the robustness of the exposure-guidance method. Since the HDR-VDP score of RB+EGB is the highest when $\alpha=3.92\%$, $3.92\%$ is selected as the threshold.

\subsubsection{Choice of Weight Function}
\begin{table*}[htbp]
	\setlength{\abovecaptionskip}{0pt}
	\setlength{\belowcaptionskip}{0pt}
	\centering
	\scriptsize
	\caption{Quantitative comparison baseline and RB+EGB with three different weights. The best results are shown in bold, and the second-best results are shown in blue. The results come from the VETHDR-Nikon test set.
		\label{tab:weight_exp}
	}
	\resizebox{2.0\columnwidth}{!}{\begin{tabular}{c|ccccccccc}
			\toprule[1pt]
			\textbf{Method} & \textbf{HDR-MAE} & \textbf{HDR-MSE} & \textbf{HDR-VDP}& \textbf{HDR-PSNR-RGB}  & \textbf{HDR-SSIM-RGB}  & \textbf{HDR-PSNR-Y}  & \textbf{HDR-SSIM-Y}  & \textbf{P [$10^{6}$]} & \textbf{FLOPs [$10^{11}$]}\\
			\midrule
			\textbf{Baseline} & {2.982} & {53.874} & {65.59} & {41.38} & {0.9786} & {43.99} & {0.9860} & \textbf{1.221} & \textbf{2.813}\\
			\midrule
			\textbf{Robertson's function \cite{robertson}} & {2.926} & \textbf{42.686} & \textbf{65.86} & {41.66} & {\color{blue}0.9791} & {44.26} & {\color{blue}0.9863} &  {\color{blue}1.850} & {\color{blue}4.263}\\
			\textbf{Debevec's function \cite{debevec1997}} & {\color{blue}2.879} & {\color{blue}45.644} & {\color{blue}65.85} & {\color{blue}41.67} & {0.9790} & \textbf{44.35} & {0.9863} & {\color{blue}1.850} & {\color{blue}4.263}\\
			\midrule
			\textbf{Our function} & \textbf{2.861} & {47.835} & {65.83} & \textbf{41.72} & \textbf{0.9794} & {\color{blue}44.32} & \textbf{0.9865} & {\color{blue}1.850} & {\color{blue}4.263}\\
			\bottomrule[1pt]
	\end{tabular}}
\end{table*}

Equation (\ref{eq:mask}) is compared with two commonly used weight functions, Debevec's function \cite{debevec1997} and Robertson's function \cite{robertson}. The results of the baseline and RB+EGB with three different weight functions are shown in Table \ref{tab:weight_exp}. The RB+EGB improves significantly from the baseline no matter what weight functions are adopted. This demonstrates that the gains are mainly coming from the design of EGB. Compared with the weight functions in \cite{debevec1997,robertson}, the Equation (\ref{eq:mask})
achieves comparable performances. The simple and concise Equation (\ref{eq:mask}) is selected as the final weight function.

\subsubsection{Working Mechanism}

The learned parameter $\beta$ in Fig. \ref{fig:overall} is reported in Table \ref{tab:beta}.
All values are between 0.3757 and 0.9734, which means that the EGB provides useful information for the restoration of full-resolution HDR images. The $\beta$ tends to increase as the network deepens, indicating that the compensated information from the EGB is important in a deep level.

\begin{figure*}[htb]
	\setlength{\abovecaptionskip}{0pt}
	\setlength{\belowcaptionskip}{0pt}
	\begin{center}
		\includegraphics[width=1\linewidth]{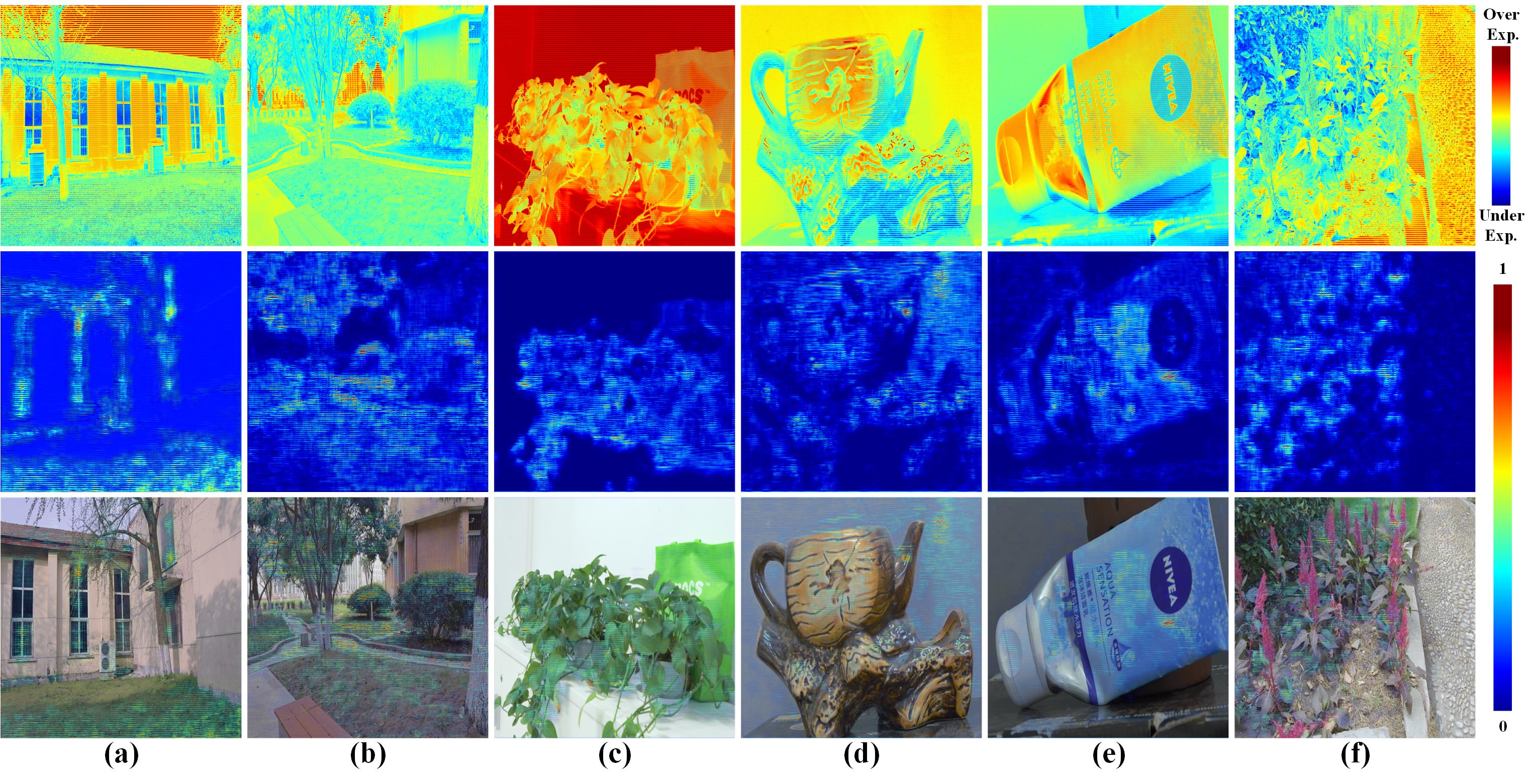}
	\end{center}
	\caption{Jet color map. The first row shows the input dual-time Bayer radiance image. The second row shows the the output feature from the last layer of EGB. The third row shows the overlay of the EGB's feature output and tone-mapped ground truth. The results come from the VETHDR-Nikon test set.}
	\label{fig:EGB_fea6}
\end{figure*}

The EGB's output is also visualized. Sixty-four output feature maps are clustered into one feature map through principal component analysis (PCA). Then, the clustered feature map is normalized into 0 to 1 and visulized using a jet color map. as shown in the second row of Fig. \ref{fig:EGB_fea6}. 

The first row shows the input dual-time Bayer radiance image, where the horizontal stripes are caused by ill exposing. The third row demonstrates the overlay of the EGB's output and tone-mapped ground truth. The EGB's features are concentrated on the well-exposed areas.  For the ill-exposed areas, e.g., the sky in column (a) and the building in the distance in colomn (b), few activations occur. Therefore, the EGB can focus on extracting features from the well-exposed areas to compensate for the main branch, which avoids the effect of saturation noise caused by ill exposing.

More visualization examples are shown in the last column of Fig. \ref{fig:EGB_exp}. All the other methods lead to unnatural color distortions. On the contrary, the compensation information from the EGB enables accurate HDR restoration. This result proves that the compensation information can effectively improve the imaging quality of the RB.

\subsection{Ablation Study}
\label{ablation}

Experiments on different models are conducted to validate the necessity of each part in our proposed framework. 
We use an RB with 16 blocks as the baseline and compare 3 models with it. When the SVC is not used, it is replaced with Opt-base. The first one has SVC in the network without EGB. The second one has EGB in the network without SVC. 
The third one is our proposed complete model, which has two SVCs and one EGB, the SVC is in front of the RB and EGB respectively.

\begin{table*}[htbp]
	
	\setlength{\abovecaptionskip}{0pt}
	\setlength{\belowcaptionskip}{0pt}
	\centering
	\scriptsize
	\caption{Quantitative comparison of models with different components. The best results are shown in bold, and the second-best results are shown in blue. The results come from the VETHDR-Nikon test set.
		\label{tab:Ablation_jpg}
	}
	\resizebox{2.0\columnwidth}{!}{\begin{tabular}{c|ccccccccc}
			\toprule[1pt]
			\textbf{Method} & \textbf{HDR-MAE} & \textbf{HDR-MSE} & \textbf{HDR-VDP}& \textbf{HDR-PSNR-RGB}  & \textbf{HDR-SSIM-RGB}  & \textbf{HDR-PSNR-Y}  & \textbf{HDR-SSIM-Y}  & \textbf{P [$10^{6}$]} & \textbf{FLOPs [$10^{11}$]}\\
			\midrule
			RB & {2.982} & {53.874} & {65.59} & {41.38} & {0.9786} & {43.99} & {0.9860} & \textbf{1.221} & \textbf{2.813}\\
			RB+SVC & {2.881} & {\color{blue}40.052} & {65.69} & {\color{blue}41.91} & {0.9791} & {\color{blue}44.33} & {0.9861} & {\color{blue}1.234} & {\color{blue}2.816}\\
			RB+EGB & {\color{blue}2.861} & {47.835} & {\color{blue}65.83} & {41.72} & {\color{blue}0.9794} & {44.32} & {\color{blue}0.9865} & {1.850} & {4.263}\\
			\midrule
			RB+2xSVC+EGB & \textbf{2.777} & \textbf{38.713} & \textbf{66.02} & \textbf{42.15} & \textbf{0.9797} & \textbf{44.56} & \textbf{0.9865} & {1.912} & {4.352}\\
			\bottomrule[1pt]
	\end{tabular}}
\end{table*}

\begin{table*}[htbp]
	
	\setlength{\abovecaptionskip}{0pt}
	\setlength{\belowcaptionskip}{0pt}
	\centering
	\scriptsize
	\caption{Quantitative comparison of models with different components. The best results are shown in bold, and the second-best results are shown in blue. The results come from the VETHDR-Canon test set.
		\label{tab:Ablation_raw}
	}
	\resizebox{2.0\columnwidth}{!}{\begin{tabular}{c|ccccccccc}
			\toprule[1pt]
			\textbf{Method} & \textbf{HDR-MAE} & \textbf{HDR-MSE} & \textbf{HDR-VDP}& \textbf{HDR-PSNR-RGB}  & \textbf{HDR-SSIM-RGB}  & \textbf{HDR-PSNR-Y}  & \textbf{HDR-SSIM-Y}  & \textbf{P [$10^{6}$]} & \textbf{FLOPs [$10^{11}$]}\\
			\midrule
			RB & {2.290} & {26.957} & {70.41} & {42.01} & {0.9889} & {43.70} & {0.9922} & \textbf{1.221} & \textbf{2.813}\\
			RB+SVC & {2.280} & {25.408} & {70.43} & {42.20} & {0.9890} & {43.75} & {0.9921} & {\color{blue}1.234} & {\color{blue}2.816}\\
			RB+EGB & {\color{blue}2.151} & {\color{blue}24.382} & \textbf{70.67} & {\color{blue}42.47} & {\color{blue}0.9895} & {\color{blue}44.06} & {\color{blue}0.9925} & {1.850} & {4.263}\\
			\midrule
			RB+2xSVC+EGB & \textbf{2.146} & \textbf{23.572} & {\color{blue}70.64} & \textbf{42.57} & \textbf{0.9897} & \textbf{44.12} & \textbf{0.9926} & {1.912} & {4.352}\\
			\bottomrule[1pt]
	\end{tabular}}
\end{table*}

The test results are shown in Table \ref{tab:Ablation_jpg} and Table \ref{tab:Ablation_raw}. They present that RB+SVC and RB+EGB have significant improvements concerning the RB from the 7 evaluation metrics points of view, which demonstrates the effectiveness of the proposed SVC and EGB. In terms of the complete model, RB+2xSVC+EGB surpasses others in image quality.

\begin{figure*}[htb] 
	\centering
	\setlength{\belowcaptionskip}{-0.2cm}
	\begin{subfigure}[b]{0.24\linewidth}
		\centering
		\includegraphics[width=40mm]{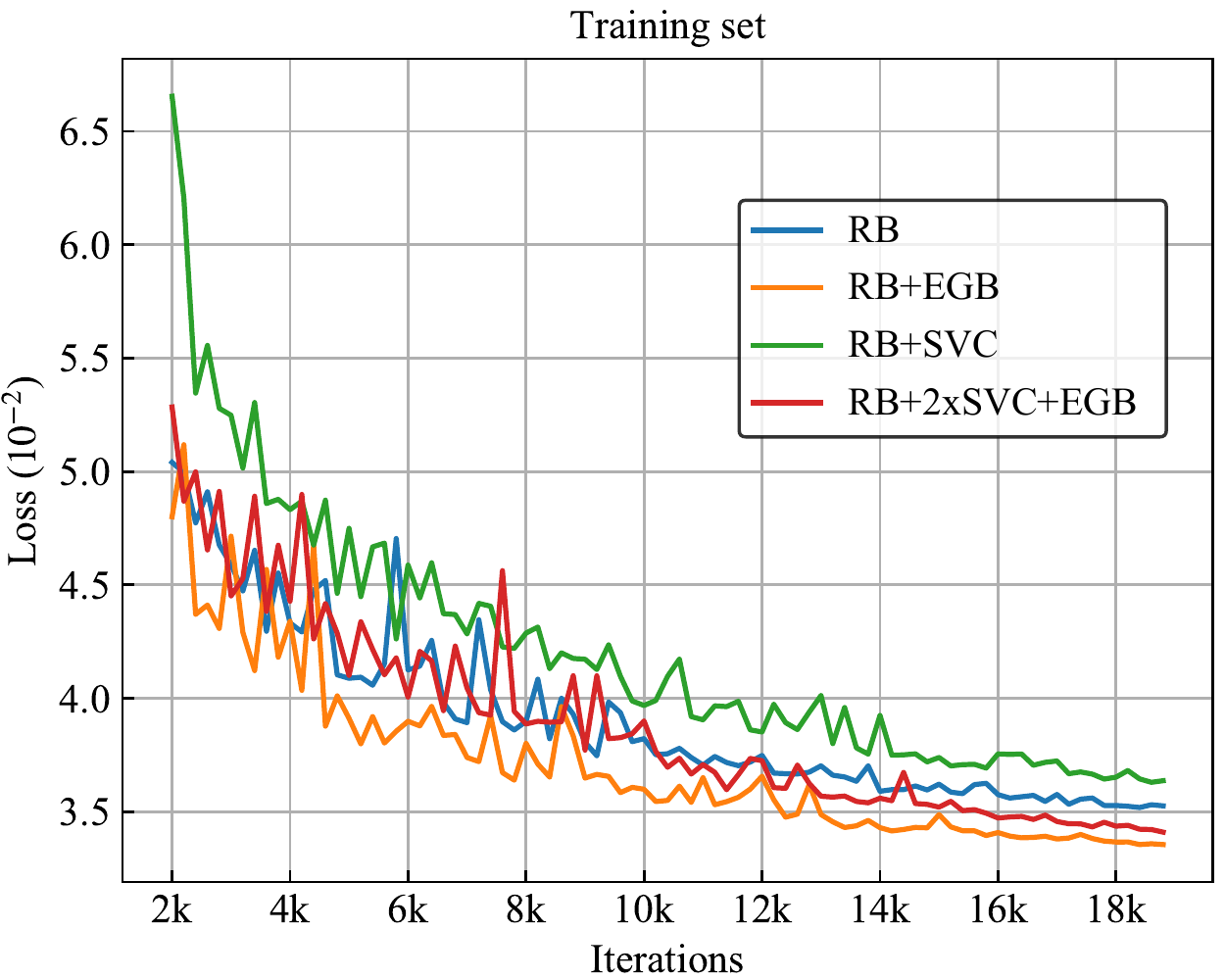}
	\end{subfigure} %
	\begin{subfigure}[b]{0.24\linewidth}
		\centering
		\includegraphics[width=40mm]{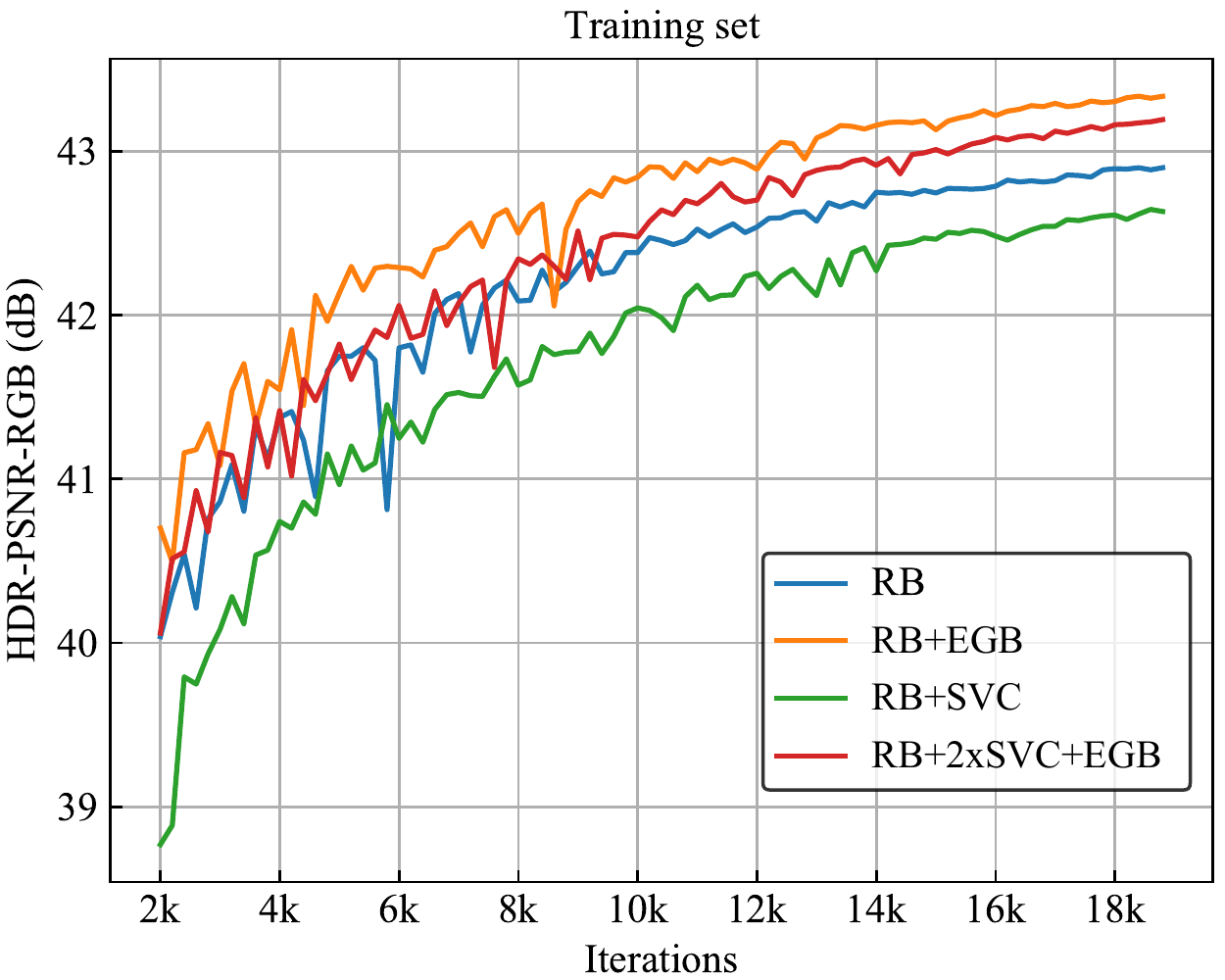}
	\end{subfigure} %
	\begin{subfigure}[b]{0.24\linewidth}
		\centering
		\includegraphics[width=40mm]{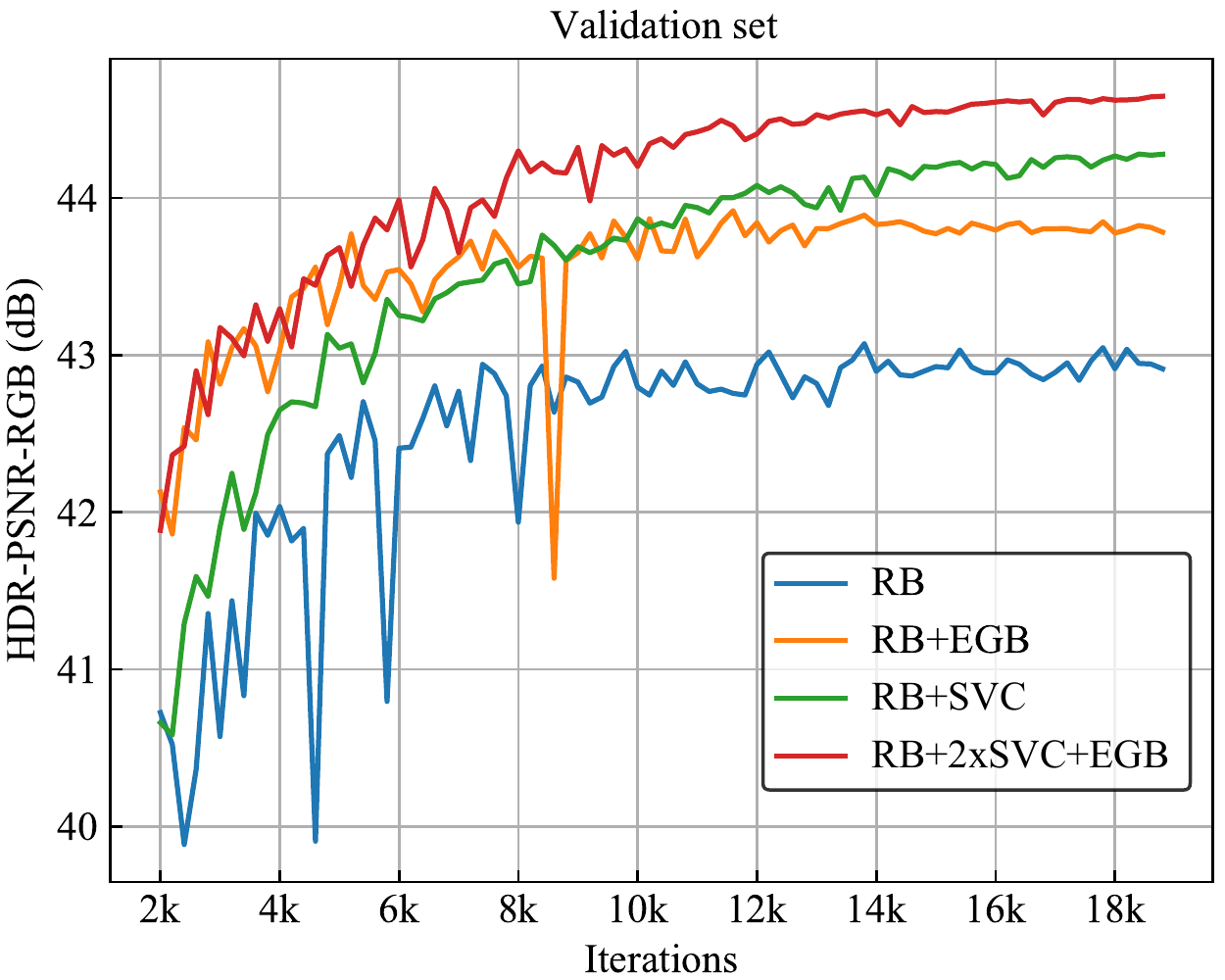}
	\end{subfigure} %
	\begin{subfigure}[b]{0.24\linewidth}
		\centering
		\includegraphics[width=40mm]{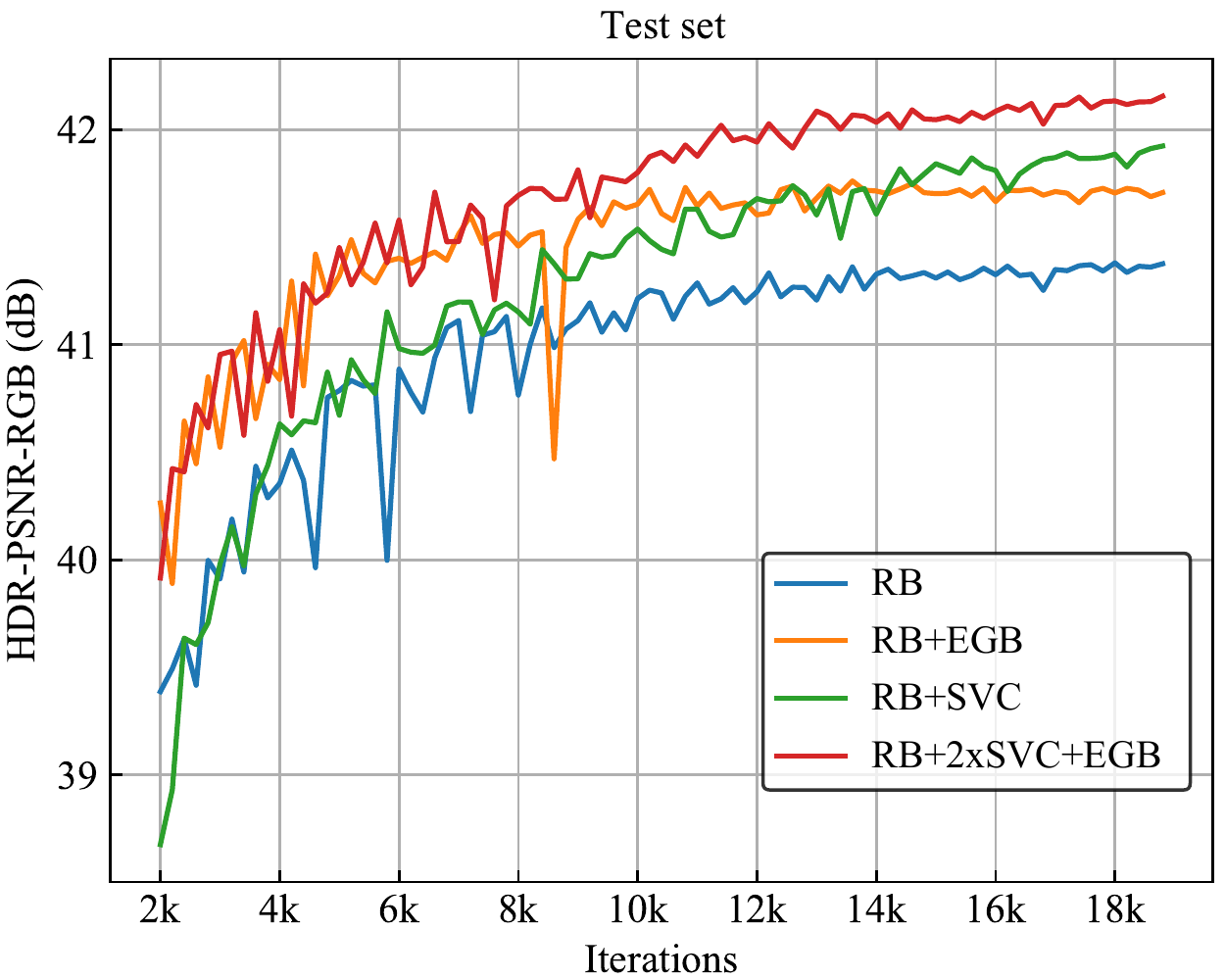}
	\end{subfigure} %
	\caption{The training plot on the training set, validation set, and test set. The results come from the VETHDR-Nikon test set.}
	\label{fig:plot-jpg}
\end{figure*}

\begin{figure*}[htb] 
	\centering
	\setlength{\belowcaptionskip}{-0.2cm}
	\begin{subfigure}[b]{0.24\linewidth}
		\centering
		\includegraphics[width=40mm]{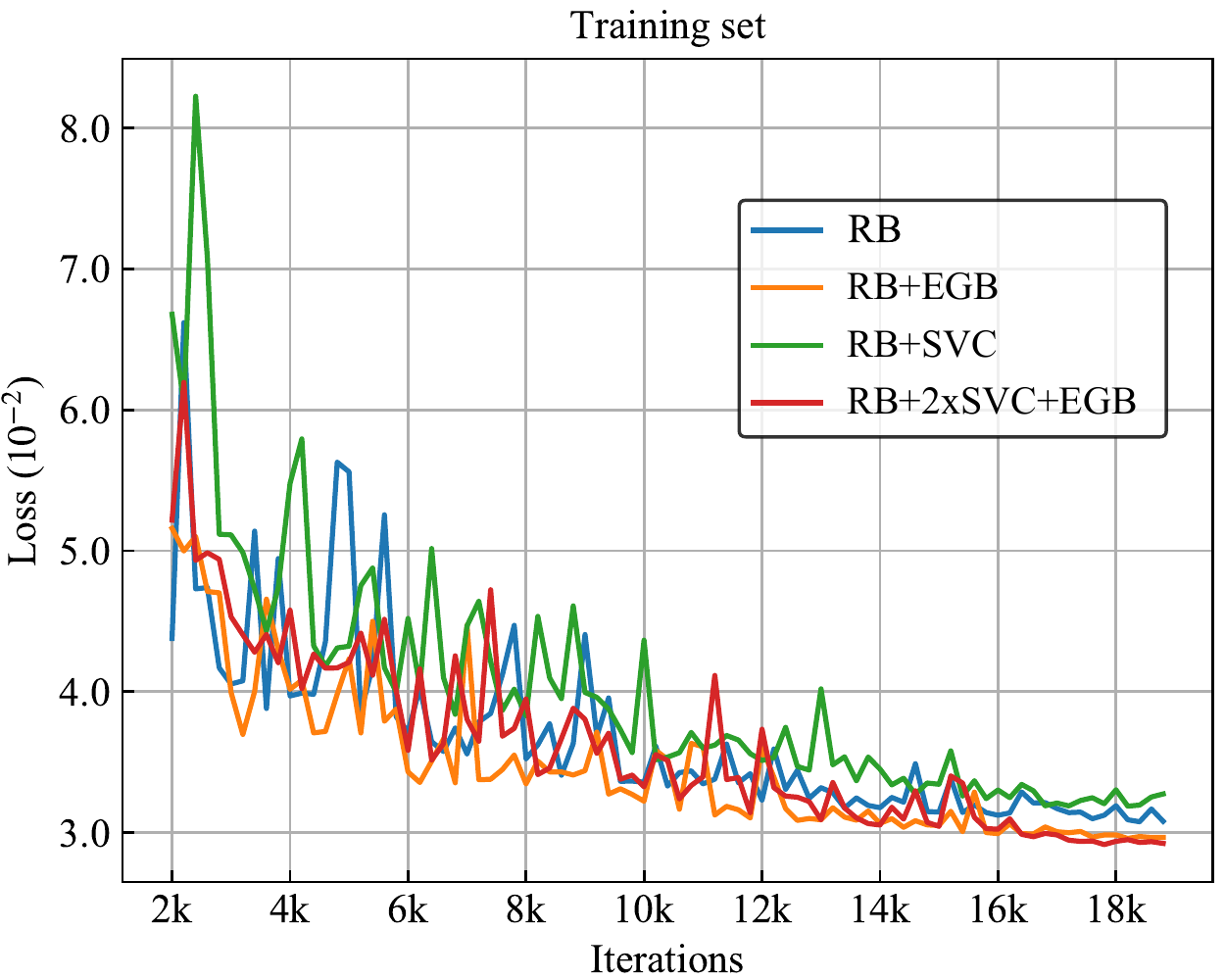}
	\end{subfigure} %
	\begin{subfigure}[b]{0.24\linewidth}
		\centering
		\includegraphics[width=40mm]{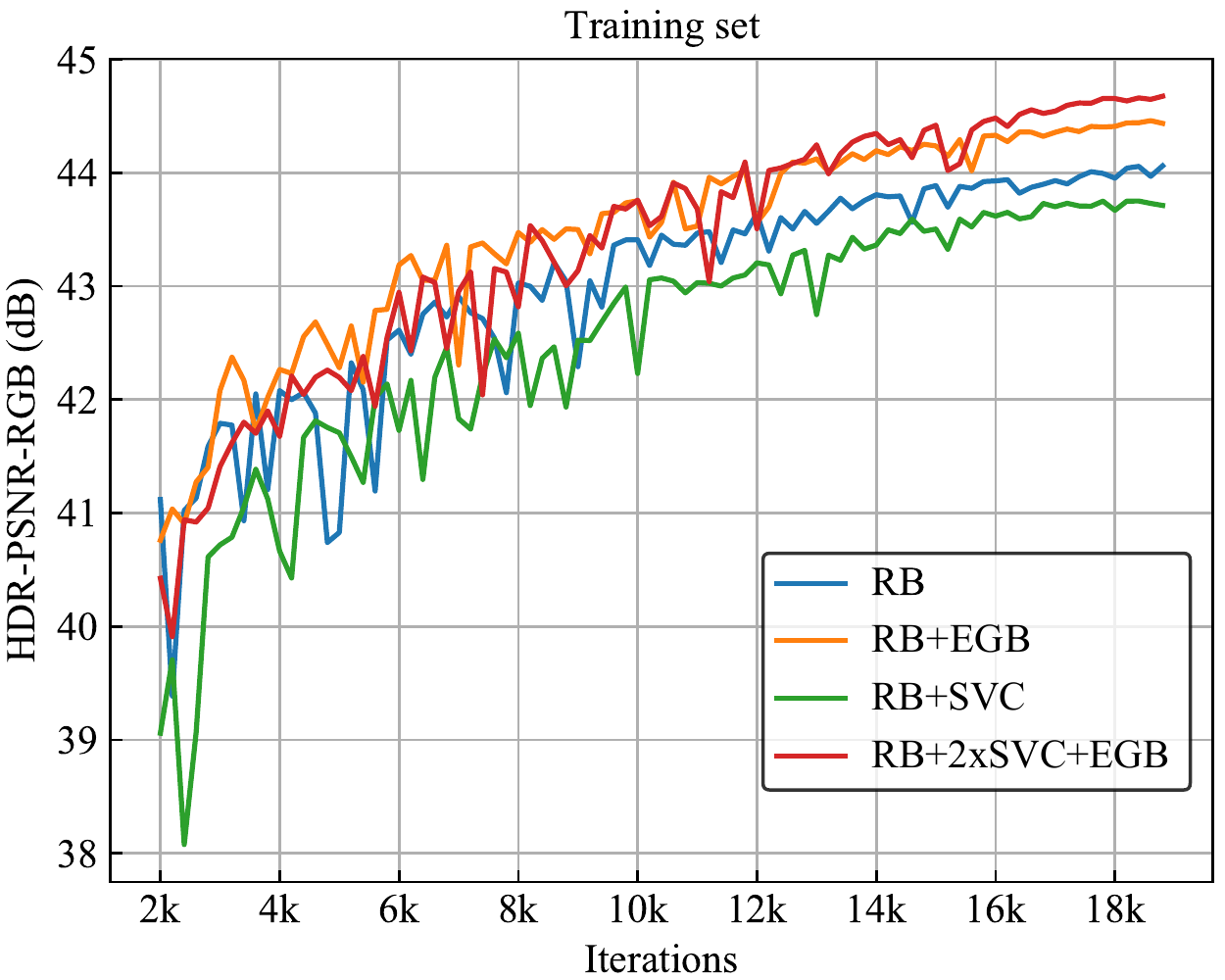}
	\end{subfigure} %
	\begin{subfigure}[b]{0.24\linewidth}
		\centering
		\includegraphics[width=40mm]{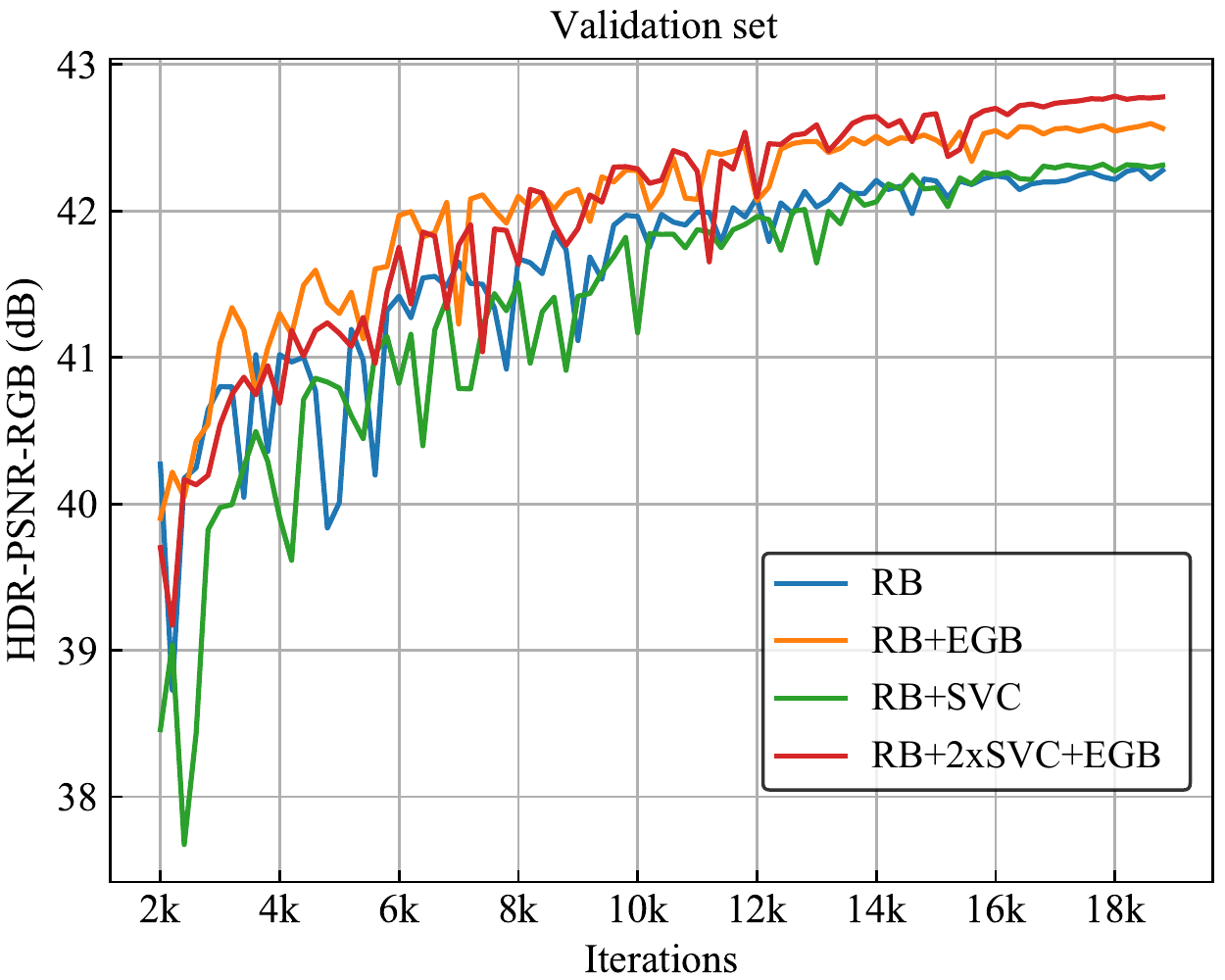}
	\end{subfigure} %
	\begin{subfigure}[b]{0.24\linewidth}
		\centering
		\includegraphics[width=40mm]{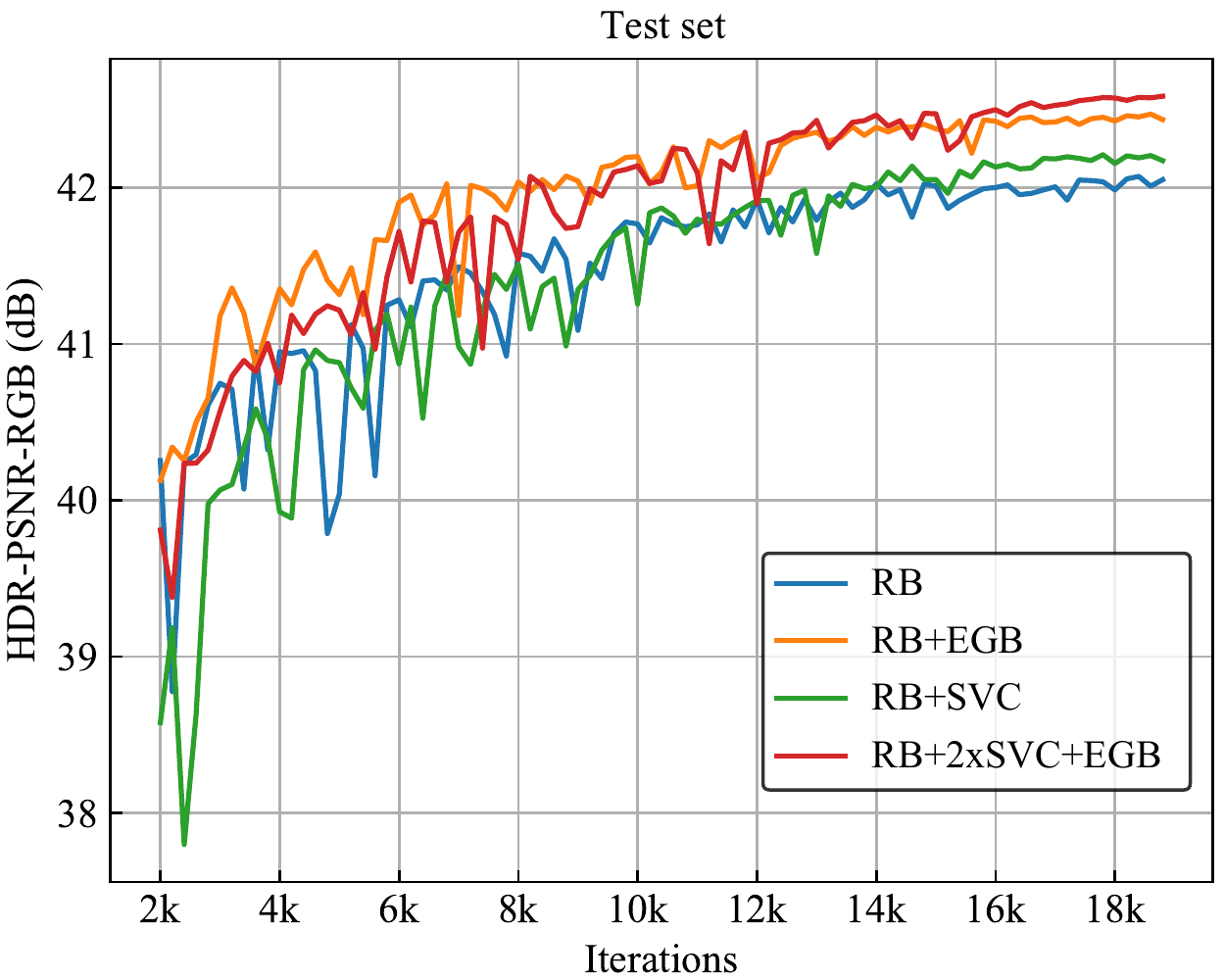}
	\end{subfigure} %
	\caption{The training plot on the training set, validation set, and test set. The results come from the VETHDR-Canon dataset.}
	\label{fig:plot-raw}
\end{figure*}

In order to verify the effectiveness of SVC and EGB more comprehensively, the training plots of RB, RB+SVC, RB+EGB, and RB+2xSVC+EGB are shown in Fig. \ref{fig:plot-jpg} and Fig. \ref{fig:plot-raw}. We can see that though RB has better convergence on training sets, it is not as robust as RB+SVC and RB+EGB on validation and test sets. Thus, both SVC and EGB improve the generalization ability of CNN. In addition, RB+2xSVC+EGB performs the best across both validation and test sets.

\subsection{Speed Evaluation}

\begin{table}[htbp]
	\setlength{\abovecaptionskip}{0pt}
	\setlength{\belowcaptionskip}{0pt}
	\centering
	\scriptsize
	\caption{The FLOPs and running time required to process images at different resolutions with RB+2xSVC+EGB.
		\label{tab:running_time}
	}
	\begin{tabular}{ccc}
		\toprule[1pt]
		\textbf{Input Resolution} & \textbf{FLOPs [$10^{11}$]} & \textbf{Time(ms)} \\
		\midrule
		 $120 \times 120$ & 0.272 & 17.943  \\
		 $240 \times 240$ & 1.088 & 71.772  \\
		 $360 \times 360$ & 2.448 & 161.486  \\
		 $480 \times 480$ & 4.352 & 287.086  \\
		 $600 \times 600$ & 6.800 & 448.573  \\
		 $720 \times 720$ & 9.793 & 645.945 \\
		 $840 \times 840$ & 13.329 & 879.202 \\
		 $960 \times 960$ & 17.409 & 1148.346 \\
		\bottomrule[1pt]
	\end{tabular}
\end{table}

In order to evaluate the running speed of our complete model RB+2xSVC+EGB when dealing with different resolutions, Table \ref{tab:running_time} lists the test results of the RB+2xSVC+EGB on GTX 1080Ti. Note that the running time of this model are proportional to the FLOPs and image resolution.

\subsection{Comparison with Existing Algorithms}

\begin{table*}[htbp]
	\setlength{\abovecaptionskip}{0pt}
	\setlength{\belowcaptionskip}{0pt}
	\centering
	\scriptsize
	\caption{Quantitative comparison between existing methods and our complete model. The best results are shown in bold, and the second-best results are shown in blue. The results come from the VETHDR-Nikon test set.
		\label{tab:final_exp_jpg}
	}
	\resizebox{2.0\columnwidth}{!}{\begin{tabular}{c|ccccccccc}
			\toprule[1pt]
			\textbf{Method} & \textbf{HDR-MAE} & \textbf{HDR-MSE} & \textbf{HDR-VDP}& \textbf{HDR-PSNR-RGB}  & \textbf{HDR-SSIM-RGB}  & \textbf{HDR-PSNR-Y}  & \textbf{HDR-SSIM-Y}  & \textbf{P [$10^{6}$]} & \textbf{FLOPs [$10^{11}$]} \\
			\midrule 
			Gharbi \textit{et al.}  \cite{dm2016} & {3.699} & {56.522} & {64.47} & {39.79} & {0.9703} & {42.06} & {0.9794} & \textbf{0.561} & \textbf{0.330}\\
			Xu \textit{et al.}  \cite{SRRI} & {3.191} & {48.28} & {65.31} & {40.89} & {0.9759} & {43.31} & {0.9841} & {\color{blue}1.021} & {\color{blue}0.467}\\
			An \textit{et al.} \cite{LeeASC}& {4.167} & {71.635} & {61.64} & {38.95} & {0.9659} & {41.43} & {0.9776} & {9.480} & {4.616}\\
			An \textit{et al.} \cite{LeeAccess}& {6.050} & {92.197} & {63.84} & {37.20} & {0.9676} & {38.95} & {0.9781} & {53.631} & {13.196}\\
			Akyuz \textit{et al.} \cite{dualiso-TIP}& {3.414} & {61.856} & {61.90} & {39.79} & {0.9692} & {42.46} & {0.9803} & {1.553} & {3.579}\\
			Suda \textit{et al.} \cite{ACCV}& {6.707} & {334.793} & {56.60} & {34.83} & {0.9428} & {37.84} & {0.9596} & {----} & {----} \\
			Hajisharif \textit{et al.} \cite{dualiso-C++}& {23.745} & {1446.367} & {56.58} & {25.93} & {0.8305} & {27.24} & {0.8584} & {----} & {----}\\
			Serrano \textit{et al.} \cite{cscode}& {49.531} & {4041.573} & {56.70} & {22.99} & {0.9239} & {23.59} & {0.9326} & {----} & {----}\\
			Xu \textit{et al.} \cite{ICME2021} & {\color{blue}2.855} & {\color{blue}40.88} & {\color{blue}65.81} & {\color{blue}41.89} & {\color{blue}0.9790} & {\color{blue}44.32} & {\color{blue}0.9861} & {2.073} & {4.773}\\
			\midrule
			Ours & \textbf{2.777} & \textbf{38.713} & \textbf{66.02} & \textbf{42.15} & \textbf{0.9797} & \textbf{44.56} & \textbf{0.9865} & {1.912} & {4.352}\\
			\bottomrule[1pt]
	\end{tabular}}
\end{table*}

\begin{table*}[htbp]
	\setlength{\abovecaptionskip}{0pt}
	\setlength{\belowcaptionskip}{0pt}
	\centering
	\scriptsize
	\caption{Quantitative comparison between existing methods and our complete model. The best results are shown in bold, and the second-best results are shown in blue. The results come from the VETHDR-Canon test set.
		\label{tab:final_exp_raw}
	}
	\resizebox{2.0\columnwidth}{!}{\begin{tabular}{c|ccccccccc}
			\toprule[1pt]
			\textbf{Method} & \textbf{HDR-MAE} & \textbf{HDR-MSE} & \textbf{HDR-VDP}& \textbf{HDR-PSNR-RGB}  & \textbf{HDR-SSIM-RGB}  & \textbf{HDR-PSNR-Y}  & \textbf{HDR-SSIM-Y}  & \textbf{P [$10^{6}$]} & \textbf{FLOPs [$10^{11}$]}\\
			\midrule 
			Gharbi \textit{et al.}  \cite{dm2016} & {2.529} & {36.102} & {69.77} & {40.81} & {0.9864} & {42.04} & {0.9899} & \textbf{0.561} & \textbf{0.330}\\
			Xu \textit{et al.}  \cite{SRRI} & {\color{blue}2.310} & {28.142} & {70.37} & {41.83} & {0.9886} & {43.31} & {0.9918} & {\color{blue}1.021} & {\color{blue}0.467}\\
			An \textit{et al.} \cite{LeeASC}& {3.774} & {60.669} & {57.79} & {38.10} & {0.9744} & {39.80} & {0.9813} & {9.480} & {4.616}\\
			An \textit{et al.} \cite{LeeAccess}& {5.547} & {59.729} & {69.31} & {37.76} & {0.9856} & {38.30} & {0.9893} & {53.631} & {13.196}\\
			Akyuz \textit{et al.} \cite{dualiso-TIP}& {3.353} & {58.502} & {57.78} & {38.35} & {0.9754} & {40.27} & {0.9823} & {1.553} & {3.579}\\
			Suda \textit{et al.} \cite{ACCV}& {9.732} & {269.536} & {57.27} & {31.85} & {0.9392} & {34.23} & {0.9481} & {----} & {----}\\
			Hajisharif \textit{et al.} \cite{dualiso-C++}& {10.982} & {416.847} & {61.07} & {29.72} & {0.8542} & {31.32} & {0.8980} & {----} & {----}\\
			Serrano \textit{et al.} \cite{cscode}& {69.954} & {7468.299} & {52.54} & {19.06} & {0.8725} & {19.44} & {0.8841} & {----} & {----}\\
			Xu \textit{et al.} \cite{ICME2021} & {2.366} & {\color{blue}25.157} & {\color{blue}70.59} & {\color{blue}42.23} & {\color{blue}0.9895} & {\color{blue}43.78} & {\color{blue}0.9925} & {2.073} & {4.773}\\
			\midrule
			Ours & \textbf{2.146} & \textbf{23.572} & \textbf{70.64} & \textbf{42.57} & \textbf{0.9897} & \textbf{44.12} & \textbf{0.9926} & {1.912} & {4.352}\\
			\bottomrule[1pt]
	\end{tabular}}
\end{table*}

The proposed algorithm is compared with seven single-shot HDRI algorithm \cite{LeeASC,LeeAccess,dualiso-TIP,ACCV,dualiso-C++,cscode,ICME2021} and two one-stage joint demosaicing algorithms \cite{dm2016,SRRI} qualitatively and quantitatively on the two datasets. The denoising and super-resolution in \cite{dm2016,SRRI} are disabled because they are not required by the single-shot dual-time HDRI.

The results about 7 evaluation metrics are reported in Table \ref{tab:final_exp_jpg} and Table \ref{tab:final_exp_raw}. With the fouth least number of parameters (P), our method achieves the best performance in HDR-MAE, HDR-MSE, HDR-VDP, HDR-PSNR-RGB, HDR-SSIM-RGB, HDR-PSNR-Y, and HDR-SSIM-Y on the test set. This superiority is mainly benefited from the proposed SVC and EGB for the single-shot HDRI.

\begin{figure*}[htb]
	\setlength{\abovecaptionskip}{0pt}
	\setlength{\belowcaptionskip}{0pt}
	\begin{center}
		\includegraphics[width=1\linewidth]{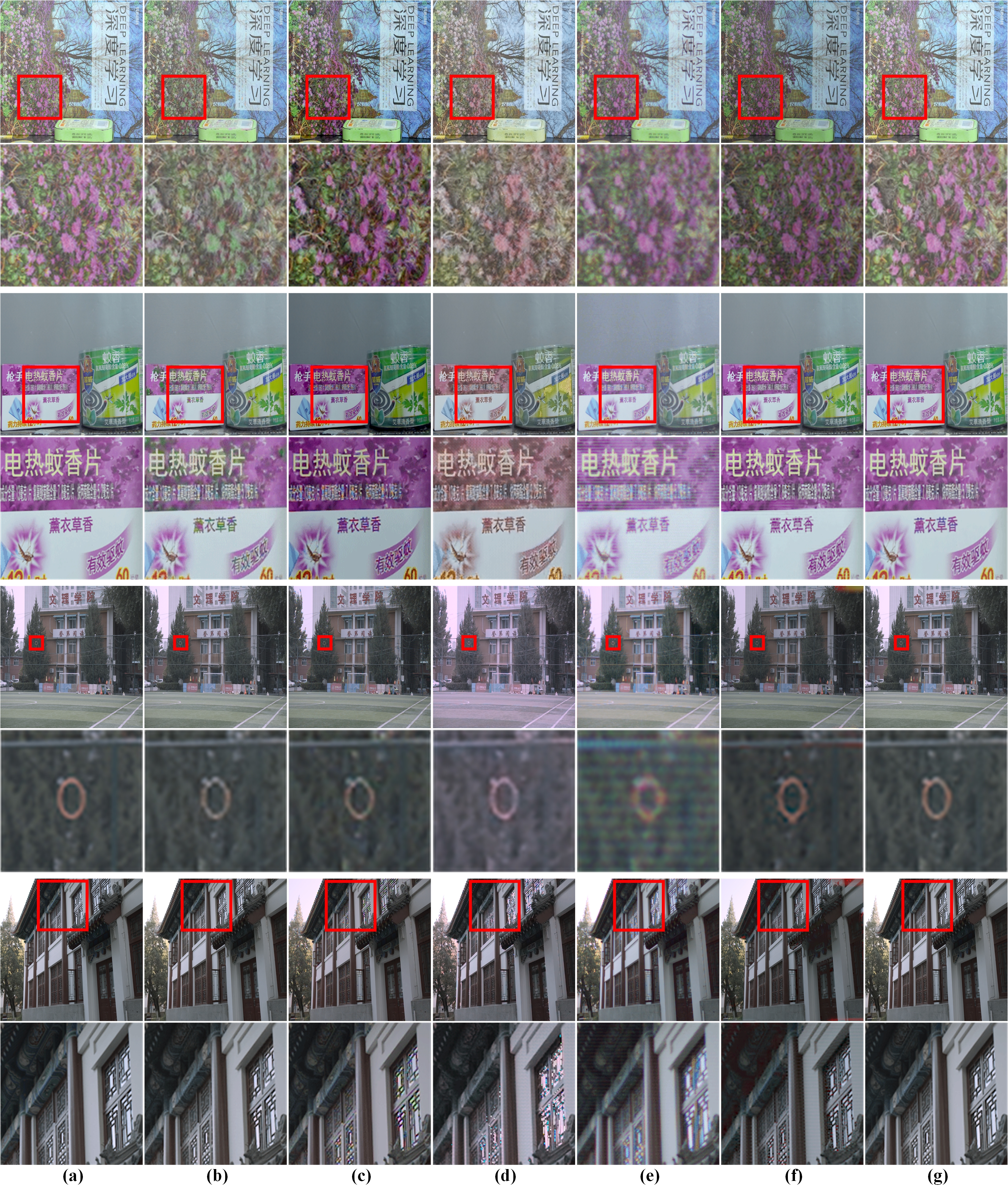}
	\end{center}
	\caption{Qualitative comparison between existing methods and our complete model. (a) Ground truth, 
			(b) An \textit{et al.}  \cite{LeeAccess}, (c) Akyuz \textit{et al.}  \cite{dualiso-TIP}, (d) Suda \textit{et al.}  \cite{ACCV}, (e) Hajisharif \textit{et al.} \cite{dualiso-C++}, (f) Serrano \textit{et al.} \cite{cscode}, (g) Ours. The results come from the VETHDR-Nikon test set and VETHDR-Canon test set.}
	\label{fig:final_exp}
\end{figure*}

For a qualitative comparison, the tone-mapping algorithm \cite{tonemap} is applied to compress the generated radiance maps for display. The synthesized results and their detailed parts are shown in Fig. \ref{fig:final_exp}. In the first row, the color of flowers is synthesized incorrectly in \textbf{(b)} and \textbf{(d)}. The texture of the flower is blurred in \textbf{(e)}. The images in \textbf{(c)} and \textbf{(f)} are generally slightly darker than the ground truth. In the second row, the luminance of images in \textbf{(c)} and \textbf{(e)} is different from the ground-truth, and the color of images in \textbf{(b)} and \textbf{(d)} is abnormal. Evident horizontal stripes can also be observed from the results in \textbf{(e)} and \textbf{(f)}.

In a nutshell, the results in \textbf{(e)} and \textbf{(f)} are likely to have horizontal stripes, which are caused by the brightness differences in the dual-time Bayer image. The results in \textbf{(c)} tend to have an inappropriate luminance, whereas the results in \textbf{(b)} and \textbf{(d)} have unnatural color. Different from these methods, the proposed algorithm is robust to most scenes, providing accurate HDR restoration. The overall comparison demonstrates the effectiveness of the proposed EGB and SVC.

\section{Conclusion Remarks and Discussion}
\label{conclusion}
In this paper, a novel CNN-based method is proposed to restore high-quality HDR images at full resolution for single-shot HDRI. The proposed CNN includes two distinctive components, the SVC and the exposure-guidance method by which the CNN is more explainable. Experimental results demonstrate that the proposed algorithm outperforms a few existing algorithms. The proposed algorithm focuses on HDR images. The idea of joint demosaicing and HDRI within a single shot can avoid cumulative errors. Since single-shot HDRI has the advantage of not having to take multiple shots, the proposed algorithm can also be extended for HDR videos \cite{HDR-video-SVE}. 

Note that the two distinctive components can be extended to other low-level image processing tasks. The SVC can be easily inserted into other networks to process Bayer image with or without SVE, and can be redesigned into flexible variants when the data pattern or SVE changes. The exposure-guidance method can be used to study other HDRI problems, such as \cite{chaobing,1zheng2021}, allowing CNN to reduce the interference of ill-exposed pixels. It is worth noting that the proposed algorithm can be further improved through the related technologies of augmentation and transfer learning. All these problems will be studied in our future research.

\section*{Acknowledgment}
This work is supported by the National Natural Science
Foundation of China under the research project 61620106012.

\ifCLASSOPTIONcaptionsoff
\newpage
\fi

\balance
\bibliographystyle{IEEEtran}
\bibliography{refs}

\end{document}